\documentclass[aps, prl, reprint, superscriptaddress]{revtex4-2}

\usepackage{graphicx}
\usepackage{dcolumn}
\usepackage{bm}

\usepackage{graphicx}
\usepackage{multirow}

\preprint{APS/123-QED}

\begin{document}

\title{\textbf{A light-induced charge order mode in a metastable cuprate ladder}}%

\author{Hari Padma}
\email{hpadmanabhan@g.harvard.edu}
\affiliation{Department of Physics, Harvard University, Cambridge, MA, USA}

\author{Prakash Sharma}
\affiliation{Department of Chemistry, Emory University, Atlanta, GA, USA}

\author{Sophia F. R. TenHuisen}
\affiliation{Department of Physics, Harvard University, Cambridge, MA, USA}
\affiliation{Department of Applied Physics, Harvard University, Cambridge, MA, USA}

\author{Filippo Glerean}
\email{Present address: Condensed Matter Physics and Materials Science Department, Brookhaven National Laboratory, Upton, NY, USA}
\affiliation{Department of Physics, Harvard University, Cambridge, MA, USA}

\author{Antoine Roll}
\affiliation{Condensed Matter Physics and Materials Science Department, Brookhaven National Laboratory, Upton, NY, USA}

\author{Pan Zhou}
\affiliation{Department of Chemistry, Emory University, Atlanta, GA, USA}

\author{Sarbajaya Kundu}
\affiliation{Department of Physics and Astronomy, University of Notre Dame, Notre Dame, IN, USA}
\affiliation{Stavropoulos Center for Complex Quantum Matter, University of Notre Dame, Notre Dame, IN, USA}

\author{Arnau Romaguera}
\author{Elizabeth Skoropata}
\author{Hiroki Ueda}
\author{Biaolong Liu}
\author{Eugenio Paris}
\affiliation{PSI Center for Photon Science, Paul Scherrer Institute, Villigen, Switzerland}

\author{Yu Wang}
\author{Seng Huat Lee}
\author{Zhiqiang Mao}
\affiliation{Department of Physics, Pennsylvania State University, University Park, PA, USA}
\affiliation{2D Crystal Consortium, Materials Research Institute, Pennsylvania State University, University Park, PA, USA}

\author{Mark P. M. Dean}
\affiliation{Condensed Matter Physics and Materials Science Department, Brookhaven National Laboratory, Upton, NY, USA}

\author{Edwin W. Huang}
\affiliation{Department of Physics and Astronomy, University of Notre Dame, Notre Dame, IN, USA}
\affiliation{Stavropoulos Center for Complex Quantum Matter, University of Notre Dame, Notre Dame, IN, USA}

\author{Elia Razzoli}
\affiliation{PSI Center for Photon Science, Paul Scherrer Institute, Villigen, Switzerland}

\author{Yao Wang}
\affiliation{Department of Chemistry, Emory University, Atlanta, GA, USA}

\author{Matteo Mitrano}
\email{mmitrano@g.harvard.edu}
\affiliation{Department of Physics, Harvard University, Cambridge, MA, USA}

\begin{abstract}
We report the observation of an emergent charge order mode in the optically-excited cuprate ladder Sr$_{14}$Cu$_{24}$O$_{41}$. Near-infrared light in the ladder plane drives a symmetry-protected electronic metastable state together with a partial melting of the equilibrium charge order. Our time-resolved resonant inelastic x-ray scattering measurements at the upper Hubbard band reveal a gapless collective excitation dispersing from the charge-order wavevector up to 0.8 eV with a slope on the order of the quasiparticle velocity. These findings reveal a regime where correlated carriers acquire itinerant character at finite momentum, and charge order becomes dynamically fluctuating, offering a platform to explore light-induced pairing instabilities.
\end{abstract}

\maketitle

Optically excited quantum materials display striking nonequilibrium phenomena, ranging from photoinduced magnetic~
\cite{disa2020polarizing,shin2018phonon} and charge-ordered phases~
\cite{Kogar2020light} to transient topological~
\cite{wang2013observation,mciver2020light,Ito2023buildup} and superconducting states~
\cite{fausti2011light,hu2014optically,mitrano2016possible}. Yet, these driven states are often short-lived, rapidly relaxing back to equilibrium due to dissipation and decoherence. In rare cases, light can steer materials into nonequilibrium configurations that evade thermalization and acquire metastable character. Such long-lived states generally arise from cooperative structural and electronic effects, defect dynamics, or trapping by impurities~
\cite{vonderlinde1974multiphoton,Koshihara1990photoinduced,Kiryukhin1997xray,fiebig1998visualization,rini2007control,Ichikawa2011transient,zhang2016cooperative,Stojchevska2015ultrafast,stoica2019optical,Nova2019metastable,disa2023photo,Vogelgesang2018phase,zong2019evidence,Gerasimenko2019quantum,budden2021evidence}. Metastability, however, may also emerge from purely electronic dynamical bottlenecks, and realizing electronic long-lived, or ``hidden”, states is a key challenge in the study of dynamical phase transitions.

Strongly correlated materials provide a natural platform for realizing electronic metastability~\cite{murakami2025photoinduced}. Theoretical works have proposed the creation of light-induced states protected against rapid decay by the presence of a Mott gap \cite{kollath2007quench, eckstein2009thermalization}, by approximate conservation laws \cite{kollar2011generalized}, or by transient trapping in the presence of competing orders~\cite{sun2020transient,Masoumi2025metastability}. Electronic nonthermal phases may also be engineered by transiently modifying the electronic Hamiltonian, which involves coherently renormalizing electronic interactions via optical dressing \cite{oka2019floquet} (either transiently or in a steady state) and enabling persistent changes in carrier distributions. These nonequilibrium states are predicted to exhibit novel electronic phases including excitonic order \cite{kunevs2015excitonic, werner2020nonthermal} and $\eta$-pairing condensates \cite{yang1989eta, rosch2008metastable, li2020eta}. However, long-lived nonequilibrium electronic states have remained experimentally elusive, hindering the exploration of novel charge correlations.

\begin{figure}
    \centering
    \includegraphics[width=0.48\textwidth]{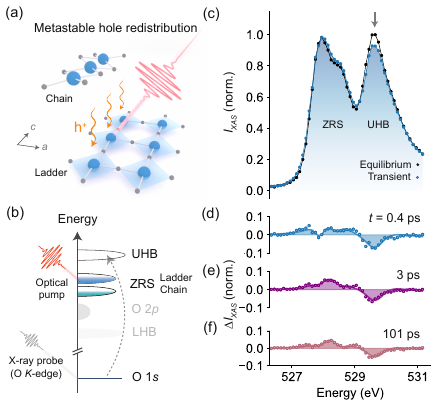} 
    \caption{(a) Crystal structure of Sr$_{14}$Cu$_{24}$O$_{41}$ featuring chain and ladder sublattices. Effectively independent at equilibrium, these units become transiently coupled by near-infrared (1.55 eV) excitation, giving rise to a metastable chain-to-ladder hole transfer. (b) Time-resolved x-ray absorption spectroscopy (trXAS) at the O $K$-edge probes the low energy electronic structure of chains and ladders, including upper Hubbard band (UHB) and Zhang-Rice singlet (ZRS) states. (c) Equilibrium (black) and transient (blue, 8 mJ/cm$^2$ fluence) O $K$-edge trXAS at $t = 0.4$ ps. Pump and probe pulses are polarized along the ladder legs ($E \parallel c$). (d-f) Differential trXAS intensity [$I_\mathrm{XAS}(t) - I_\mathrm{XAS}(t < 0)$] at $t = 0.4$ ps, 3 ps, and 101 ps, respectively. The pump-induced spectral reshaping is metastable. The solid lines are fits to the data (see SM Section 1).}
    \label{fig:fig1}
\end{figure}

Recent experiments have observed electronic metastability in the model cuprate ladder Sr$_{14}$Cu$_{24}$O$_{41}$ \cite{padma2025symmetry}. Starting from a self-doped ($p$ = 0.06) charge-ordered ground state \cite{abbamonte2004crystallization,Vuletic2006spinladder}, ultrafast 1.55-eV pump pulses induce a prompt transfer of holes between the chain charge reservoir and ladder subunits, which are effectively decoupled at equilibrium [see Fig. \ref{fig:fig1}(a)]. The ladder hole concentration increases ($\Delta p$ = 0.03), partially melting charge order and driving the system toward a gapless regime. This nonequilibrium charge redistribution arises from a coherent optical dressing of Zhang–Rice singlet states in the ladders and the activation of a symmetry-forbidden chain-to-ladder hopping term. When the optical field vanishes, the equilibrium symmetry is restored, suppressing relaxation and trapping holes in a nonthermal state persisting for several nanoseconds. Beyond demonstrating symmetry-protected electronic metastability, these results prompt the question of whether trapped carriers exhibit emergent collective dynamics or incipient pairing, as expected from the intrinsic tendency of cuprate ladders toward hole binding and superconductivity under pressure \cite{uehara1996superconductivity, dagotto1999experiments, hirthe2023magnetically, padma2025beyond, scheie2025cooper}.

Here, we explore this phenomenon by probing the charge dynamics of driven Sr$_{14}$Cu$_{24}$O$_{41}$ using O $K$-edge time-resolved resonant inelastic x-ray scattering (trRIXS). Following optical excitation into the metastable state, we observe a novel collective mode emerging from the partially melted charge order. This excitation disperses up to 0.8~eV from the ordering wavevector with a slope on the order of the quasiparticle velocity, as expected for a charge collective mode. Based on its energy scale and dispersion, and supported by numerical simulations, we attribute this mode to charge order fluctuations in the ladders. These measurements indicate that the electronic metastable state hosts gapless, fluctuating charge order dynamics distinct from those of charge-ordered cuprates at equilibrium, and where correlated carriers exhibit itinerant character at finite momentum.

We conducted time-resolved x-ray absorption (trXAS) and trRIXS measurements at the Furka endstation of the Athos beamline at SwissFEL, Paul Scherrer Institut \cite{swissfel_furka}. We acquired trXAS spectra in fluorescence yield mode with x-rays at near-normal incidence, detecting the signal with an avalanche photodiode positioned at $2\theta = 78^\mathrm{o}$. We recorded O $K$-edge trRIXS spectra with an incident energy of 530 eV, keeping the scattering angle fixed at $2\theta = 136^\mathrm{o}$ while varying the incident angle $\theta$ from 70$^\mathrm{o}$ to 130$^\mathrm{o}$. This geometry corresponds to momentum transfers from 0 to 0.28 r.l.u. (in units of 2$\pi$/$c_\mathrm{L}$, $c_\mathrm{L}$ = 3.95 Å) along the ladder legs and accesses the tail of the charge-order peak at $q_\mathrm{CO}$ = (0, 1, 0.2) r.l.u. The RIXS spectrometer provided a total energy resolution of 160 meV. We used $\sigma$-polarized x-rays to enhance the intensity of charge excitations and focused the beam to 300 $\mu$m ($H$) $\times$ 10 $\mu$m ($V$). We monitored shot-to-shot x-ray intensity fluctuations with a photodiode and used them to normalize the data. We excited the samples with 800 nm (1.55 eV), 100-fs pulses focused to 500 $\mu$m to achieve fluences up to 8 mJ/cm$^2$, as in our prior experiments \cite{padma2025symmetry}. The pump penetration depth exceeds that of the soft x-rays at all measured incident angles, ensuring a homogeneously excited volume. We cleaved high-quality single crystals of Sr$_{14}$Cu$_{24}$O$_{41}$ in situ along the $ac$ plane and maintained them at 100 K throughout the experiment.

Resonant x-ray scattering at the O $K$-edge ($1s\!\rightarrow\!2p$ transition) of copper oxides provides a direct probe of hole ordering and collective excitations. The x-ray absorption spectrum of Sr$_{14}$Cu$_{24}$O$_{41}$ in this energy range exhibits two prominent pre-edge features at 528 and 529.7~eV, corresponding to Zhang-Rice singlet states (ZRS) and upper Hubbard band (UHB), respectively [Fig.~\ref{fig:fig1}(b)]. The relative intensities of these peaks encode the distribution of holes between chains and ladders, as established in equilibrium studies with Ca substitution~\cite{nucker2000hole, abbamonte2004crystallization}. Following optical excitation, the trXAS spectra [Fig.~\ref{fig:fig1}(c)] display a prompt spectral-weight transfer from the UHB to the ZRS peak, which remains largely unchanged for hundreds of picoseconds [Figs.~\ref{fig:fig1}(d)–(f)]. This reshaping is consistent with our previous report of metastable chain-to-ladder hole transfer~\cite{padma2025symmetry} and with the equilibrium behavior of Ca-substituted compounds~\cite{nucker2000hole}. Crucially, these spectral changes reveal that correlated carriers in the UHB undergo the most pronounced dynamical changes in the metastable state.

\begin{figure*} 
    \includegraphics[width=0.8\textwidth]{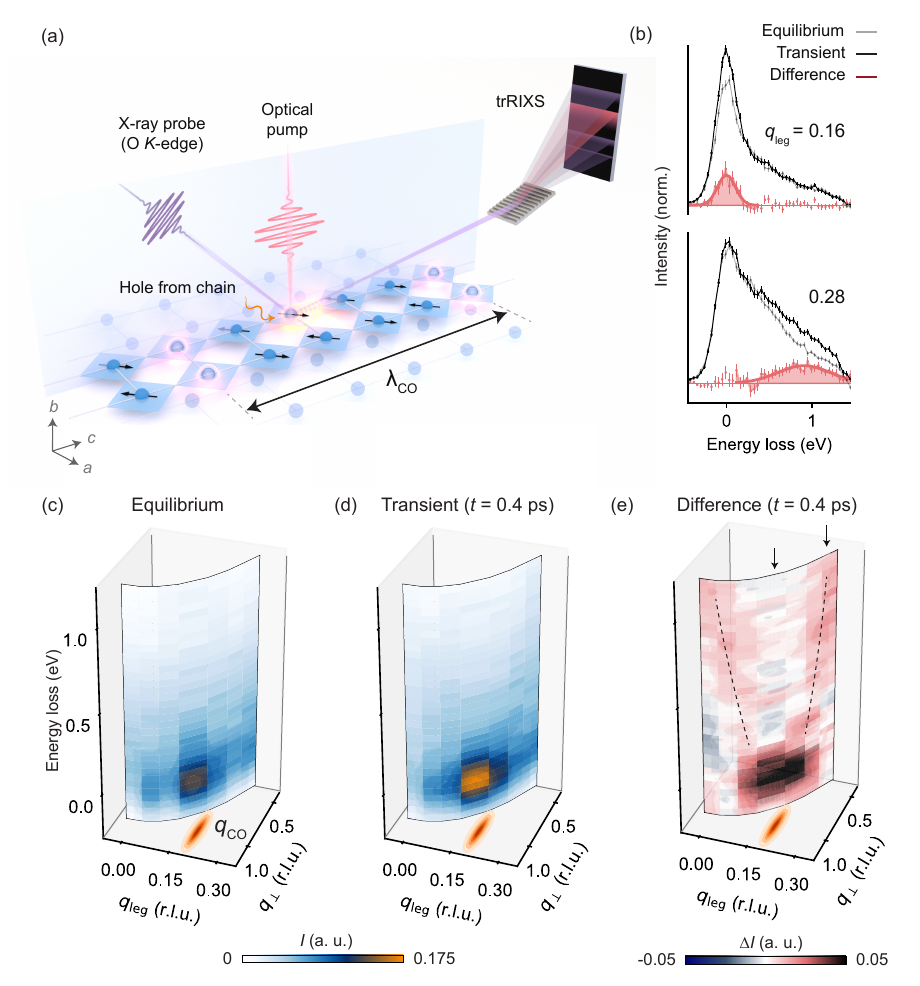} 
    \caption{(a) Sketch of the time-resolved resonant inelastic X-ray scattering (trRIXS) experiment. Following optical excitation, O $K$-edge X-ray pulses probe charge dynamics in the ladder via scattering into a grating spectrometer. Black arrows represent spins, pink halos represent charge-ordered holes in the ground state, and the yellow halo a hole transferred from the chain to the ladder. $\lambda_\mathrm{CO}$ denotes the charge-order wavelength. (b) Equilibrium (grey), transient (black), and difference (red) trRIXS spectra at $t$ = 0.4 ps, at momentum transfers $q_\mathrm{leg}$ = 0.16 and 0.28 reciprocal lattice units (r.l.u.) and at resonance with the UHB peak of the XAS spectrum. Here, $q_\mathrm{leg} \parallel c$ and $q_\perp \parallel b$. (c-e) Equilibrium, transient, and difference trRIXS momentum-energy intensity maps, as a function of $q_\mathrm{leg}$ and $q_{\perp}$, respectively.  $q_\mathrm{CO}$ marks the charge order wavevector \cite{abbamonte2004crystallization}. Arrows in (e) indicate the momentum transfer of the spectra in panel (b). Dashed lines in (e) are guides to the eye.}
    \label{fig:fig2}
\end{figure*}

We probe the collective excitations of these correlated carriers using high-resolution trRIXS~\cite{mitrano2020probing,Mitrano2024exploring}. Fig.~\ref{fig:fig2} summarizes the main findings of this Letter. To suppress fluorescence that overlaps with the RIXS signal and directly probe the dynamics of the UHB states, we tune the incident x-rays to the 529.7-eV resonance. With the scattering geometry set in the $ac$ plane, we measure the dispersion along the ladder legs ($q_\mathrm{leg}$) while simultaneously intercepting the tail of the charge-order peak $q_\mathrm{CO}$~\cite{abbamonte2004crystallization,Rusydi2006quantum} [Fig.~\ref{fig:fig2}(a)]. Although charge order is most prominent at resonance with the Zhang–Rice peak, it is still visible at the UHB resonance. Upon excitation into the metastable phase, the trRIXS spectra exhibit a pronounced momentum-dependent reshaping at energy losses below 1.2~eV [Fig.~\ref{fig:fig2}(b)]. At $q_\mathrm{leg} = 0.16$ r.l.u., the quasielastic peak is enhanced while the finite-energy-loss spectrum remains unchanged. In contrast, at $q_\mathrm{leg} = 0.28$ r.l.u., the quasielastic response is largely unaltered, and a strong enhancement appears at energy losses up to 1.2~eV. These momentum-dependent spectral changes signal the emergence of a new collective excitation.

We construct energy–momentum maps of the trRIXS spectra to resolve this emergent excitation. We rotate the sample angle while keeping the x-ray analyzer fixed, thereby scanning the momentum transfer along the ladder direction. Figs.~\ref{fig:fig2}(c–e) display the resulting spectra as a function of $q_\mathrm{leg}$ (parallel to the ladder legs) and $q_\perp$ (normal to the ladder planes). The momentum trajectory intercepts the tail of the charge-order peak at $q_\mathrm{CO}$, which contributes to the quasielastic intensity. The equilibrium map [Fig.~\ref{fig:fig2}(c)] shows the corresponding enhancement of the quasielastic peak near the tail of the charge-order reflection, together with a broad continuum of excitations up to $\sim$0.6~eV, attributed to the $\Delta S = 0$ two-triplon continuum~\cite{tseng2023momentum}. In the metastable phase [Fig.~\ref{fig:fig2}(d)], the transient RIXS signal exhibits an enhancement of spectral weight up to 1.2~eV, accompanied by an increased quasielastic charge-order peak. This enhancement is consistent with diffuse scattering around the charge order peak and a partial redistribution of spectral weight from the elastic peak to its tails. To highlight these changes, we plot the difference between transient and equilibrium maps in Fig.~\ref{fig:fig2}(e), revealing a well-defined and strongly dispersive mode centered at the tail of $q_\mathrm{CO}$. 

\begin{figure*}
    \includegraphics[width=0.75\textwidth]{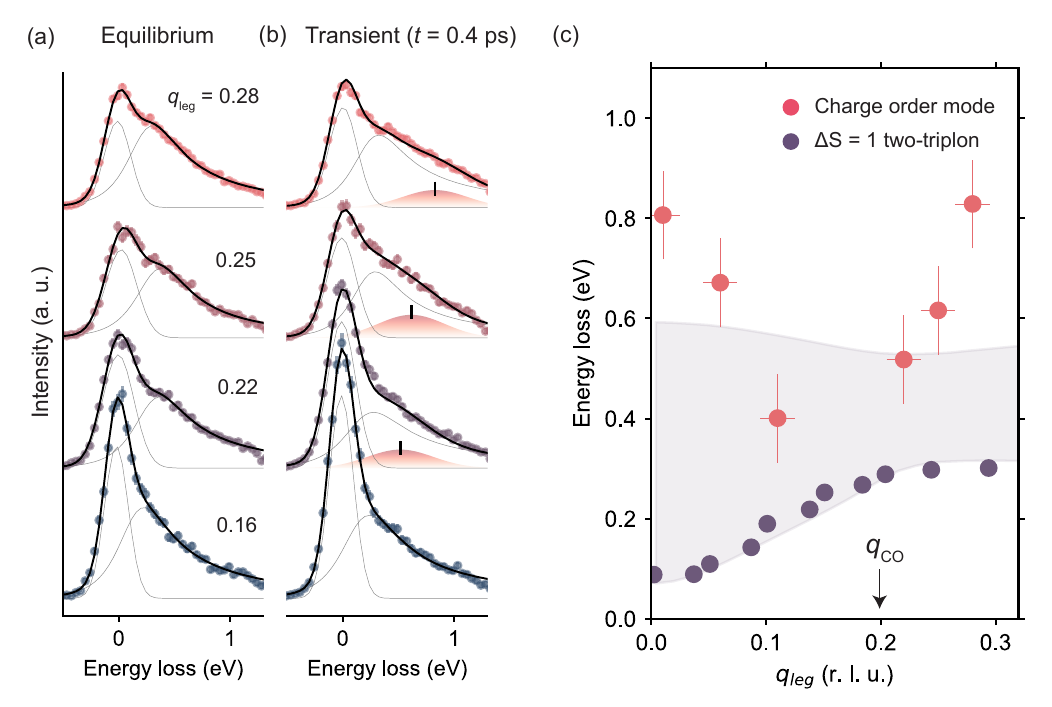} 
    \caption{(a,b) Momentum-dependent RIXS spectra in and out of equilibrium, vertically offset for clarity. Momentum is given in r.l.u. along the ladder direction. Equilibrium spectra in (a) are fit with a Gaussian for the quasielastic peak and an asymmetric Lorentzian for the $\Delta S = 0$ two-triplon (thin black lines). Transient spectra in (b) include an additional Gaussian component (shaded red). (c) Dispersion of the emergent collective mode (red markers) extracted from fits in (b). The shaded gray area denotes the theoretical $\Delta S = 0$ two-triplon continuum from Ref.~\cite{tseng2023momentum}, while blue markers indicate the lower boundary of the $\Delta S = 1$ two-triplon continuum from Ref.~\cite{padma2025beyond}. Error bars represent instrumental uncertainties in momentum transfer and energy-loss reference point.}
    \label{fig:fig3}
\end{figure*}

We quantify the mode dispersion by fitting the momentum-dependent trRIXS spectra. Our prior Cu $L$-edge trRIXS experiments showed that the metastable hole transfer disrupts ladder spin singlets, suppressing the $\Delta S = 1$ two-triplon continuum intensity without any measurable energy shift and indicating that the exchange coupling constants remain unchanged~\cite{padma2025symmetry}. This observation constrains the O $K$-edge trRIXS analysis. We first model the equilibrium spectra with a Gaussian for the quasielastic peak and an asymmetric Lorentzian for the $\Delta S = 0$ two-triplon continuum [Fig.~\ref{fig:fig3}(a)]. For the transient spectra, we use the equilibrium parameters as initial conditions and include an additional Gaussian to capture the high-energy enhancement [Fig.~\ref{fig:fig3}(b)]. The resulting dispersion of the emergent mode is shown in Fig.~\ref{fig:fig3}(c). These fits reproduce the observed spectral redistribution without any systematic shift of the magnetic continuum. In the metastable state, the intensity of the $\Delta S = 0$ continuum is reduced by $\sim$10\%, consistent with the suppression of the $\Delta S = 1$ spin-excitation intensity at the Cu $L$ edge~\cite{padma2025symmetry}. In contrast, fitting the transient spectra without the additional term requires an overall blueshift of the $\Delta S = 0$ continuum to reproduce the high-energy enhancement, at odds with our prior measurements~\cite{padma2025symmetry}. The energy scale and dispersion of the light-induced mode are also incompatible with a magnetic origin. Its propagation velocity (2.1~$\pm$~0.3~eV\AA) far exceeds that of $\Delta S = 1$ magnetic excitations in this ladder compound~\cite{padma2025beyond} (lower boundary shown in Fig.~\ref{fig:fig3}(c)) and those observed in other cuprates. Moreover, it approaches energy scales larger than any magnetic excitation, including the $\Delta S = 0$ two-triplon continuum that dominates the equilibrium RIXS spectrum at the O $K$-edge UHB~\cite{tseng2023momentum}. Taken together, these observations identify the feature as a charge collective mode.

We now discuss the assignment of this emergent charge mode. This excitation has not been observed in previous equilibrium RIXS studies~\cite{tseng2023momentum}, raising the question of its origin. Its energy scale and propagation velocity resemble that of acoustic plasmons recently reported in electron- and hole-doped cuprates~\cite{lee2014asymmetry, hepting2018three, lin2020doping, nag2020detection}. However, acoustic plasmons disperse symmetrically about $q = 0$, in sharp contrast to our observation of a branch centered at the charge order wave vector $q_\mathrm{CO}$. No counterpart is detected at $q_\mathrm{leg} = 0$, arguing against a zone-folding replica as the origin of this mode. Although weak coupling theories are not necessarily accurate for this material, for completeness, we performed RPA calculations (see Fig. S6) which support this conclusion. We furthermore exclude the lower edge of the particle–hole continuum near $2k_\mathrm{F}$ as the origin. Such excitations are indeed detectable by RIXS, as shown in recent measurements \cite{lomeli2025direct}. However, $2k_\mathrm{F} = (1-p)/2$ in one-dimensional systems \cite{giamarchi2003quantum, kung2017numerically}, which for our experimental hole densities $p < 0.10$ corresponds to $q_\mathrm{leg} > 0.45$ r.l.u., well beyond the momentum range where we observe the charge mode.

The symmetric dispersion around $q_\mathrm{CO}$ is consistent with a gapless charge order collective mode. In a static charge-ordered state, the dynamical charge structure factor carries spectral weight predominantly at zero energy, whereas fluctuating order shifts spectral weight to finite energies~\cite{kivelson2003detect,mitrano2019ultrafast}. Ultrafast resonant soft x-ray diffraction reveals a partial suppression of the static charge-order peak in the metastable state~\cite{padma2025symmetry}, while our current O $K$-edge trRIXS measurements show a concurrent enhancement around $q_\mathrm{CO}$ at finite energy loss. The absence of any measurable shift in the charge-order wavevector~\cite{padma2025symmetry} rules out a sliding charge-ordered state~\cite{Mitrano2019evidence}, indicating that the melting of charge order proceeds through fluctuating, rather than translational, dynamics. The propagation velocity of the resulting excitations is comparable to that of charge-density-wave modes observed at equilibrium in two-dimensional cuprates (up to 1.3~eV\AA)~\cite{chaix2017dispersive, lee2021spectroscopic, lin2020strongly, li2020multiorbital, huang2021quantum}. Notably, it is also on the order of the ladder quasiparticle dispersion, 1.4~eV\AA, determined by angle-resolved photoemission~\cite{takahashi1997angle}, suggesting that free and periodically modulated carriers propagate in a similar manner within the metastable state.

To understand the enhanced charge fluctuations in the metastable state, we performed DMRG calculations of charge correlations in a single-band Hubbard ladder as a function of hole doping $p$ (see SM Section 4). This model has been shown to capture the spin dynamics of cuprate ladders at equilibrium~\cite{padma2025beyond, scheie2025cooper} and in the metastable state~\cite{padma2025symmetry}. Our calculations reveal a doping-driven crossover accompanied by an order-of-magnitude enhancement of charge fluctuations. For $p < 0.08$, the ground state features rapidly decaying charge correlations, indicative of highly suppressed charge fluctuations. As the hole density increases, these fluctuations grow markedly, as signaled by a sharp decrease of the power-law exponent. We identify a crossover between these regimes near $p = 0.1$ (Fig.~S7). While charge order in the parent compound Sr$_{14}$Cu$_{24}$O$_{41}$ cannot be quantitatively reproduced by the single-band Hubbard model \cite{wohlfeld2007origin, wohlfeld2010t}, our results nevertheless provide a theoretical basis to understand the charge mode observed in the metastable state~\cite{padma2025symmetry}. At equilibrium, the ladder subunits in Sr$_{14}$Cu$_{24}$O$_{41}$ are self-doped with $p = 0.06$, while in the metastable state, the photoinduced hole transfer $\Delta p = 0.03$ drives the system towards a highly fluctuating regime. These fluctuations, in the presence of the partially melted charge order, manifest as the gapless collective mode observed at $q_\mathrm{CO}$

Our experiments reveal an emergent charge order mode in the electronic metastable phase of Sr$_{14}$Cu$_{24}$O$_{41}$. Following the optically-induced transfer of holes from the chain reservoirs to the ladders, the charge-ordered state partially melts and evolves into a fluctuating phase featuring gapless charge excitations extending up to 0.8~eV. This collective mode appears as a distinct hallmark of the light-induced metastable state, absent from the equilibrium RIXS spectrum and unreported in prior studies \cite{tseng2023momentum}. 

Our findings have implications for the physics of metastable phases and dynamical ordering phenomena in correlated systems. The observation of a gapless collective mode emanating from the charge-order wavevector indicates that the charge-ordered phase enters a fluctuating regime in which correlated carriers in the UHB acquire itinerant character at finite momentum. This nonequilibrium state provides a promising platform for exploring dynamical pairing instabilities in cuprates. In two-dimensional cuprates at equilibrium, charge order competes with superconductivity, and its optical suppression leads to transient superconducting-like states~\cite{fausti2011light, nicoletti2014optically, hu2014optically,cremin2019photoenhanced}, consistent with theoretical expectations for nonequilibrium competing orders~\cite{sun2020transient, wang2021fluctuating}. Cuprate ladders display similar competition~\cite{dagotto1992superconductivity, blumberg2002sliding, vuletic2003suppression, Vuletic2006spinladder} together with strong hole pairing \cite{hirthe2023magnetically, padma2025beyond, scheie2025cooper}, suggesting that their fluctuating charge-ordered state could be driven toward a metastable superconducting or $\eta$-paired condensate under suitable conditions. Our results thus motivate future optical and x-ray scattering experiments to search for signatures of such hidden condensates~\cite{murakami2022exploring} in the spin and charge structure factors of light-driven metastable electronic states~\cite{murakami2025photoinduced}.

We thank P. Abbamonte, G. Aeppli, L. Benfatto, G. C. Ghiringhelli, and S. Johnston for insightful discussions. The experimental research leading to these results was primarily supported by the US Department of Energy, Office of Basic Energy Sciences, Early Career Award Program, under award no. DE-SC0022883. Theoretical part of the work (P.S. and Yao Wang) was supported by the Air Force Office of Scientific Research Young Investigator Program under grant no. FA9550-23-1-0153. Work performed at Brookhaven National Laboratory was supported by the US Department of Energy, Division of Materials Science, under contract no. DE-SC0012704. We acknowledge the Paul Scherrer Institut, Villigen, Switzerland, for the provision of beamtime at the Furka beamline of the SwissFEL. Work at PSI was partially funded by the PSI Research Grant 2022. The single-crystal growth work was performed at the Pennsylvania State University Two-Dimensional Crystal Consortium—Materials Innovation Platform (2DCC-MIP), which is supported by NSF Cooperative Agreement no. DMR-2039351. The simulation used resources of the Frontera computing system at the Texas Advanced Computing Center, Award No. OAC-1818253.

\bibliography{apssamp}%

\begin{thebibliography}{9}%
\makeatletter
\providecommand \@ifxundefined [1]{%
 \@ifx{#1\undefined}
}%
\providecommand \@ifnum [1]{%
 \ifnum #1\expandafter \@firstoftwo
 \else \expandafter \@secondoftwo
 \fi
}%
\providecommand \@ifx [1]{%
 \ifx #1\expandafter \@firstoftwo
 \else \expandafter \@secondoftwo
 \fi
}%
\providecommand \natexlab [1]{#1}%
\providecommand \enquote  [1]{``#1''}%
\providecommand \bibnamefont  [1]{#1}%
\providecommand \bibfnamefont [1]{#1}%
\providecommand \citenamefont [1]{#1}%
\providecommand \href@noop [0]{\@secondoftwo}%
\providecommand \href [0]{\begingroup \@sanitize@url \@href}%
\providecommand \@href[1]{\@@startlink{#1}\@@href}%
\providecommand \@@href[1]{\endgroup#1\@@endlink}%
\providecommand \@sanitize@url [0]{\catcode `\\12\catcode `\$12\catcode `\&12\catcode `\#12\catcode `\^12\catcode `\_12\catcode `\%12\relax}%
\providecommand \@@startlink[1]{}%
\providecommand \@@endlink[0]{}%
\providecommand \url  [0]{\begingroup\@sanitize@url \@url }%
\providecommand \@url [1]{\endgroup\@href {#1}{\urlprefix }}%
\providecommand \urlprefix  [0]{URL }%
\providecommand \Eprint [0]{\href }%
\providecommand \doibase [0]{https://doi.org/}%
\providecommand \selectlanguage [0]{\@gobble}%
\providecommand \bibinfo  [0]{\@secondoftwo}%
\providecommand \bibfield  [0]{\@secondoftwo}%
\providecommand \translation [1]{[#1]}%
\providecommand \BibitemOpen [0]{}%
\providecommand \bibitemStop [0]{}%
\providecommand \bibitemNoStop [0]{.\EOS\space}%
\providecommand \EOS [0]{\spacefactor3000\relax}%
\providecommand \BibitemShut  [1]{\csname bibitem#1\endcsname}%
\let\auto@bib@innerbib\@empty
\bibitem [{\citenamefont {N{\"u}cker}\ \emph {et~al.}(2000)\citenamefont {N{\"u}cker}, \citenamefont {Merz}, \citenamefont {Kuntscher}, \citenamefont {Gerhold}, \citenamefont {Schuppler}, \citenamefont {Neudert}, \citenamefont {Golden}, \citenamefont {Fink}, \citenamefont {Schild}, \citenamefont {Stadler} \emph {et~al.}}]{nucker2000hole}%
  \BibitemOpen
  \bibfield  {author} {\bibinfo {author} {\bibfnamefont {N.}~\bibnamefont {N{\"u}cker}}, \bibinfo {author} {\bibfnamefont {M.}~\bibnamefont {Merz}}, \bibinfo {author} {\bibfnamefont {C.~A.}\ \bibnamefont {Kuntscher}}, \bibinfo {author} {\bibfnamefont {S.}~\bibnamefont {Gerhold}}, \bibinfo {author} {\bibfnamefont {S.}~\bibnamefont {Schuppler}}, \bibinfo {author} {\bibfnamefont {R.}~\bibnamefont {Neudert}}, \bibinfo {author} {\bibfnamefont {M.}~\bibnamefont {Golden}}, \bibinfo {author} {\bibfnamefont {J.}~\bibnamefont {Fink}}, \bibinfo {author} {\bibfnamefont {D.}~\bibnamefont {Schild}}, \bibinfo {author} {\bibfnamefont {S.}~\bibnamefont {Stadler}}, \emph {et~al.},\ }\bibfield  {title} {\bibinfo {title} {Hole distribution in {(Sr,Ca,Y,La)}$_{14}${Cu}$_{24}${O}$_{41}$ ladder compounds studied by x-ray absorption spectroscopy},\ }\href@noop {} {\bibfield  {journal} {\bibinfo  {journal} {Physical Review B}\ }\textbf {\bibinfo {volume} {62}},\ \bibinfo {pages} {14384} (\bibinfo {year} {2000})}\BibitemShut {NoStop}%
\bibitem [{\citenamefont {Padma}\ \emph {et~al.}(2025{\natexlab{a}})\citenamefont {Padma}, \citenamefont {Glerean}, \citenamefont {TenHuisen}, \citenamefont {Shen}, \citenamefont {Wang}, \citenamefont {Xu}, \citenamefont {Elliott}, \citenamefont {Homes}, \citenamefont {Skoropata}, \citenamefont {Ueda}, \citenamefont {Liu}, \citenamefont {Paris}, \citenamefont {Romaguera}, \citenamefont {Lee}, \citenamefont {He}, \citenamefont {Wang}, \citenamefont {Lee}, \citenamefont {Choi}, \citenamefont {Park}, \citenamefont {Mao}, \citenamefont {Calandra}, \citenamefont {Jang}, \citenamefont {Razzoli}, \citenamefont {Dean}, \citenamefont {Wang},\ and\ \citenamefont {Mitrano}}]{padma2025symmetry}%
  \BibitemOpen
  \bibfield  {author} {\bibinfo {author} {\bibfnamefont {H.}~\bibnamefont {Padma}}, \bibinfo {author} {\bibfnamefont {F.}~\bibnamefont {Glerean}}, \bibinfo {author} {\bibfnamefont {S.~F.~R.}\ \bibnamefont {TenHuisen}}, \bibinfo {author} {\bibfnamefont {Z.}~\bibnamefont {Shen}}, \bibinfo {author} {\bibfnamefont {H.}~\bibnamefont {Wang}}, \bibinfo {author} {\bibfnamefont {L.}~\bibnamefont {Xu}}, \bibinfo {author} {\bibfnamefont {J.~D.}\ \bibnamefont {Elliott}}, \bibinfo {author} {\bibfnamefont {C.~C.}\ \bibnamefont {Homes}}, \bibinfo {author} {\bibfnamefont {E.}~\bibnamefont {Skoropata}}, \bibinfo {author} {\bibfnamefont {H.}~\bibnamefont {Ueda}}, \bibinfo {author} {\bibfnamefont {B.}~\bibnamefont {Liu}}, \bibinfo {author} {\bibfnamefont {E.}~\bibnamefont {Paris}}, \bibinfo {author} {\bibfnamefont {A.}~\bibnamefont {Romaguera}}, \bibinfo {author} {\bibfnamefont {B.}~\bibnamefont {Lee}}, \bibinfo {author} {\bibfnamefont {W.}~\bibnamefont {He}}, \bibinfo {author} {\bibfnamefont {Y.}~\bibnamefont {Wang}}, \bibinfo
  {author} {\bibfnamefont {S.~H.}\ \bibnamefont {Lee}}, \bibinfo {author} {\bibfnamefont {H.}~\bibnamefont {Choi}}, \bibinfo {author} {\bibfnamefont {S.-Y.}\ \bibnamefont {Park}}, \bibinfo {author} {\bibfnamefont {Z.}~\bibnamefont {Mao}}, \bibinfo {author} {\bibfnamefont {M.}~\bibnamefont {Calandra}}, \bibinfo {author} {\bibfnamefont {H.}~\bibnamefont {Jang}}, \bibinfo {author} {\bibfnamefont {E.}~\bibnamefont {Razzoli}}, \bibinfo {author} {\bibfnamefont {M.~P.~M.}\ \bibnamefont {Dean}}, \bibinfo {author} {\bibfnamefont {Y.}~\bibnamefont {Wang}},\ and\ \bibinfo {author} {\bibfnamefont {M.}~\bibnamefont {Mitrano}},\ }\bibfield  {title} {\bibinfo {title} {{Symmetry-protected electronic metastability in an optically driven cuprate ladder}},\ }\href {https://doi.org/10.1038/s41563-025-02254-2} {\bibfield  {journal} {\bibinfo  {journal} {Nature Materials}\ }\textbf {\bibinfo {volume} {24}},\ \bibinfo {pages} {1584} (\bibinfo {year} {2025}{\natexlab{a}})}\BibitemShut {NoStop}%
\bibitem [{\citenamefont {Bertinshaw}\ \emph {et~al.}(2020)\citenamefont {Bertinshaw}, \citenamefont {Kim}, \citenamefont {Porras}, \citenamefont {Ueda}, \citenamefont {Sung}, \citenamefont {Efimenko}, \citenamefont {Bombardi}, \citenamefont {Kim}, \citenamefont {Keimer},\ and\ \citenamefont {Kim}}]{bertinshaw2020spin}%
  \BibitemOpen
  \bibfield  {author} {\bibinfo {author} {\bibfnamefont {J.}~\bibnamefont {Bertinshaw}}, \bibinfo {author} {\bibfnamefont {J.~K.}\ \bibnamefont {Kim}}, \bibinfo {author} {\bibfnamefont {J.}~\bibnamefont {Porras}}, \bibinfo {author} {\bibfnamefont {K.}~\bibnamefont {Ueda}}, \bibinfo {author} {\bibfnamefont {N.-H.}\ \bibnamefont {Sung}}, \bibinfo {author} {\bibfnamefont {A.}~\bibnamefont {Efimenko}}, \bibinfo {author} {\bibfnamefont {A.}~\bibnamefont {Bombardi}}, \bibinfo {author} {\bibfnamefont {J.}~\bibnamefont {Kim}}, \bibinfo {author} {\bibfnamefont {B.}~\bibnamefont {Keimer}},\ and\ \bibinfo {author} {\bibfnamefont {B.~J.}\ \bibnamefont {Kim}},\ }\bibfield  {title} {\bibinfo {title} {Spin-wave gap collapse in {Rh}-doped {Sr}$_2${Ir}{O}$_4$},\ }\href@noop {} {\bibfield  {journal} {\bibinfo  {journal} {Physical Review B}\ }\textbf {\bibinfo {volume} {101}},\ \bibinfo {pages} {094428} (\bibinfo {year} {2020})}\BibitemShut {NoStop}%
\bibitem [{\citenamefont {Martinelli}\ \emph {et~al.}(2022)\citenamefont {Martinelli}, \citenamefont {Betto}, \citenamefont {Kummer}, \citenamefont {Arpaia}, \citenamefont {Braicovich}, \citenamefont {Di~Castro}, \citenamefont {Brookes}, \citenamefont {Moretti~Sala},\ and\ \citenamefont {Ghiringhelli}}]{martinelli2022fractional}%
  \BibitemOpen
  \bibfield  {author} {\bibinfo {author} {\bibfnamefont {L.}~\bibnamefont {Martinelli}}, \bibinfo {author} {\bibfnamefont {D.}~\bibnamefont {Betto}}, \bibinfo {author} {\bibfnamefont {K.}~\bibnamefont {Kummer}}, \bibinfo {author} {\bibfnamefont {R.}~\bibnamefont {Arpaia}}, \bibinfo {author} {\bibfnamefont {L.}~\bibnamefont {Braicovich}}, \bibinfo {author} {\bibfnamefont {D.}~\bibnamefont {Di~Castro}}, \bibinfo {author} {\bibfnamefont {N.~B.}\ \bibnamefont {Brookes}}, \bibinfo {author} {\bibfnamefont {M.}~\bibnamefont {Moretti~Sala}},\ and\ \bibinfo {author} {\bibfnamefont {G.}~\bibnamefont {Ghiringhelli}},\ }\bibfield  {title} {\bibinfo {title} {Fractional spin excitations in the infinite-layer cuprate {Ca}{Cu}{O}$_2$},\ }\href@noop {} {\bibfield  {journal} {\bibinfo  {journal} {Physical Review X}\ }\textbf {\bibinfo {volume} {12}},\ \bibinfo {pages} {021041} (\bibinfo {year} {2022})}\BibitemShut {NoStop}%
\bibitem [{\citenamefont {Padma}\ \emph {et~al.}(2025{\natexlab{b}})\citenamefont {Padma}, \citenamefont {Thomas}, \citenamefont {TenHuisen}, \citenamefont {He}, \citenamefont {Guan}, \citenamefont {Li}, \citenamefont {Lee}, \citenamefont {Wang}, \citenamefont {Lee}, \citenamefont {Mao}, \citenamefont {Jang}, \citenamefont {Bisogni}, \citenamefont {Pelliciari}, \citenamefont {Dean}, \citenamefont {Johnston},\ and\ \citenamefont {Mitrano}}]{padma2025beyond}%
  \BibitemOpen
  \bibfield  {author} {\bibinfo {author} {\bibfnamefont {H.}~\bibnamefont {Padma}}, \bibinfo {author} {\bibfnamefont {J.}~\bibnamefont {Thomas}}, \bibinfo {author} {\bibfnamefont {S.~F.~R.}\ \bibnamefont {TenHuisen}}, \bibinfo {author} {\bibfnamefont {W.}~\bibnamefont {He}}, \bibinfo {author} {\bibfnamefont {Z.}~\bibnamefont {Guan}}, \bibinfo {author} {\bibfnamefont {J.}~\bibnamefont {Li}}, \bibinfo {author} {\bibfnamefont {B.}~\bibnamefont {Lee}}, \bibinfo {author} {\bibfnamefont {Y.}~\bibnamefont {Wang}}, \bibinfo {author} {\bibfnamefont {S.~H.}\ \bibnamefont {Lee}}, \bibinfo {author} {\bibfnamefont {Z.}~\bibnamefont {Mao}}, \bibinfo {author} {\bibfnamefont {H.}~\bibnamefont {Jang}}, \bibinfo {author} {\bibfnamefont {V.}~\bibnamefont {Bisogni}}, \bibinfo {author} {\bibfnamefont {J.}~\bibnamefont {Pelliciari}}, \bibinfo {author} {\bibfnamefont {M.~P.~M.}\ \bibnamefont {Dean}}, \bibinfo {author} {\bibfnamefont {S.}~\bibnamefont {Johnston}},\ and\ \bibinfo {author} {\bibfnamefont {M.}~\bibnamefont {Mitrano}},\
  }\bibfield  {title} {\bibinfo {title} {{Beyond-Hubbard Pairing in a Cuprate Ladder}},\ }\href {https://doi.org/10.1103/PhysRevX.15.021049} {\bibfield  {journal} {\bibinfo  {journal} {Phys. Rev. X}\ }\textbf {\bibinfo {volume} {15}},\ \bibinfo {pages} {021049} (\bibinfo {year} {2025}{\natexlab{b}})}\BibitemShut {NoStop}%
\bibitem [{\citenamefont {Fetter}(1974)}]{fetter1974layered}%
  \BibitemOpen
  \bibfield  {author} {\bibinfo {author} {\bibfnamefont {A.~L.}\ \bibnamefont {Fetter}},\ }\bibfield  {title} {\bibinfo {title} {{Electrodynamics of a layered electron gas. II. Periodic array}},\ }\href {https://doi.org/https://doi.org/10.1016/0003-4916(74)90397-2} {\bibfield  {journal} {\bibinfo  {journal} {Annals of Physics}\ }\textbf {\bibinfo {volume} {88}},\ \bibinfo {pages} {1} (\bibinfo {year} {1974})}\BibitemShut {NoStop}%
\bibitem [{\citenamefont {Hepting}\ \emph {et~al.}(2018)\citenamefont {Hepting}, \citenamefont {Chaix}, \citenamefont {Huang}, \citenamefont {Fumagalli}, \citenamefont {Peng}, \citenamefont {Moritz}, \citenamefont {Kummer}, \citenamefont {Brookes}, \citenamefont {Lee}, \citenamefont {Hashimoto} \emph {et~al.}}]{hepting2018three}%
  \BibitemOpen
  \bibfield  {author} {\bibinfo {author} {\bibfnamefont {M.}~\bibnamefont {Hepting}}, \bibinfo {author} {\bibfnamefont {L.}~\bibnamefont {Chaix}}, \bibinfo {author} {\bibfnamefont {E.}~\bibnamefont {Huang}}, \bibinfo {author} {\bibfnamefont {R.}~\bibnamefont {Fumagalli}}, \bibinfo {author} {\bibfnamefont {Y.}~\bibnamefont {Peng}}, \bibinfo {author} {\bibfnamefont {B.}~\bibnamefont {Moritz}}, \bibinfo {author} {\bibfnamefont {K.}~\bibnamefont {Kummer}}, \bibinfo {author} {\bibfnamefont {N.}~\bibnamefont {Brookes}}, \bibinfo {author} {\bibfnamefont {W.}~\bibnamefont {Lee}}, \bibinfo {author} {\bibfnamefont {M.}~\bibnamefont {Hashimoto}}, \emph {et~al.},\ }\bibfield  {title} {\bibinfo {title} {Three-dimensional collective charge excitations in electron-doped copper oxide superconductors},\ }\href@noop {} {\bibfield  {journal} {\bibinfo  {journal} {Nature}\ }\textbf {\bibinfo {volume} {563}},\ \bibinfo {pages} {374} (\bibinfo {year} {2018})}\BibitemShut {NoStop}%
\bibitem [{\citenamefont {Chen}\ \emph {et~al.}(2021)\citenamefont {Chen}, \citenamefont {Wang}, \citenamefont {Rebec}, \citenamefont {Jia}, \citenamefont {Hashimoto}, \citenamefont {Lu}, \citenamefont {Moritz}, \citenamefont {Moore}, \citenamefont {Devereaux},\ and\ \citenamefont {Shen}}]{chen2021anomalously}%
  \BibitemOpen
  \bibfield  {author} {\bibinfo {author} {\bibfnamefont {Z.}~\bibnamefont {Chen}}, \bibinfo {author} {\bibfnamefont {Y.}~\bibnamefont {Wang}}, \bibinfo {author} {\bibfnamefont {S.~N.}\ \bibnamefont {Rebec}}, \bibinfo {author} {\bibfnamefont {T.}~\bibnamefont {Jia}}, \bibinfo {author} {\bibfnamefont {M.}~\bibnamefont {Hashimoto}}, \bibinfo {author} {\bibfnamefont {D.}~\bibnamefont {Lu}}, \bibinfo {author} {\bibfnamefont {B.}~\bibnamefont {Moritz}}, \bibinfo {author} {\bibfnamefont {R.~G.}\ \bibnamefont {Moore}}, \bibinfo {author} {\bibfnamefont {T.~P.}\ \bibnamefont {Devereaux}},\ and\ \bibinfo {author} {\bibfnamefont {Z.-X.}\ \bibnamefont {Shen}},\ }\bibfield  {title} {\bibinfo {title} {Anomalously strong near-neighbor attraction in doped {1D} cuprate chains},\ }\href@noop {} {\bibfield  {journal} {\bibinfo  {journal} {Science}\ }\textbf {\bibinfo {volume} {373}},\ \bibinfo {pages} {1235} (\bibinfo {year} {2021})}\BibitemShut {NoStop}%
\bibitem [{\citenamefont {Wang}\ \emph {et~al.}(2021)\citenamefont {Wang}, \citenamefont {Chen}, \citenamefont {Shi}, \citenamefont {Moritz}, \citenamefont {Shen},\ and\ \citenamefont {Devereaux}}]{wang2021phonon}%
  \BibitemOpen
  \bibfield  {author} {\bibinfo {author} {\bibfnamefont {Y.}~\bibnamefont {Wang}}, \bibinfo {author} {\bibfnamefont {Z.}~\bibnamefont {Chen}}, \bibinfo {author} {\bibfnamefont {T.}~\bibnamefont {Shi}}, \bibinfo {author} {\bibfnamefont {B.}~\bibnamefont {Moritz}}, \bibinfo {author} {\bibfnamefont {Z.-X.}\ \bibnamefont {Shen}},\ and\ \bibinfo {author} {\bibfnamefont {T.~P.}\ \bibnamefont {Devereaux}},\ }\bibfield  {title} {\bibinfo {title} {Phonon-mediated long-range attractive interaction in one-dimensional cuprates},\ }\href@noop {} {\bibfield  {journal} {\bibinfo  {journal} {Physical Review Letters}\ }\textbf {\bibinfo {volume} {127}},\ \bibinfo {pages} {197003} (\bibinfo {year} {2021})}\BibitemShut {NoStop}%
\end{thebibliography}%


\begin{thebibliography}{75}%
\makeatletter
\providecommand \@ifxundefined [1]{%
 \@ifx{#1\undefined}
}%
\providecommand \@ifnum [1]{%
 \ifnum #1\expandafter \@firstoftwo
 \else \expandafter \@secondoftwo
 \fi
}%
\providecommand \@ifx [1]{%
 \ifx #1\expandafter \@firstoftwo
 \else \expandafter \@secondoftwo
 \fi
}%
\providecommand \natexlab [1]{#1}%
\providecommand \enquote  [1]{``#1''}%
\providecommand \bibnamefont  [1]{#1}%
\providecommand \bibfnamefont [1]{#1}%
\providecommand \citenamefont [1]{#1}%
\providecommand \href@noop [0]{\@secondoftwo}%
\providecommand \href [0]{\begingroup \@sanitize@url \@href}%
\providecommand \@href[1]{\@@startlink{#1}\@@href}%
\providecommand \@@href[1]{\endgroup#1\@@endlink}%
\providecommand \@sanitize@url [0]{\catcode `\\12\catcode `\$12\catcode `\&12\catcode `\#12\catcode `\^12\catcode `\_12\catcode `\%12\relax}%
\providecommand \@@startlink[1]{}%
\providecommand \@@endlink[0]{}%
\providecommand \url  [0]{\begingroup\@sanitize@url \@url }%
\providecommand \@url [1]{\endgroup\@href {#1}{\urlprefix }}%
\providecommand \urlprefix  [0]{URL }%
\providecommand \Eprint [0]{\href }%
\providecommand \doibase [0]{https://doi.org/}%
\providecommand \selectlanguage [0]{\@gobble}%
\providecommand \bibinfo  [0]{\@secondoftwo}%
\providecommand \bibfield  [0]{\@secondoftwo}%
\providecommand \translation [1]{[#1]}%
\providecommand \BibitemOpen [0]{}%
\providecommand \bibitemStop [0]{}%
\providecommand \bibitemNoStop [0]{.\EOS\space}%
\providecommand \EOS [0]{\spacefactor3000\relax}%
\providecommand \BibitemShut  [1]{\csname bibitem#1\endcsname}%
\let\auto@bib@innerbib\@empty
\bibitem [{\citenamefont {Disa}\ \emph {et~al.}(2020)\citenamefont {Disa}, \citenamefont {Fechner}, \citenamefont {Nova}, \citenamefont {Liu}, \citenamefont {F{\"o}rst}, \citenamefont {Prabhakaran}, \citenamefont {Radaelli},\ and\ \citenamefont {Cavalleri}}]{disa2020polarizing}%
  \BibitemOpen
  \bibfield  {author} {\bibinfo {author} {\bibfnamefont {A.~S.}\ \bibnamefont {Disa}}, \bibinfo {author} {\bibfnamefont {M.}~\bibnamefont {Fechner}}, \bibinfo {author} {\bibfnamefont {T.~F.}\ \bibnamefont {Nova}}, \bibinfo {author} {\bibfnamefont {B.}~\bibnamefont {Liu}}, \bibinfo {author} {\bibfnamefont {M.}~\bibnamefont {F{\"o}rst}}, \bibinfo {author} {\bibfnamefont {D.}~\bibnamefont {Prabhakaran}}, \bibinfo {author} {\bibfnamefont {P.~G.}\ \bibnamefont {Radaelli}},\ and\ \bibinfo {author} {\bibfnamefont {A.}~\bibnamefont {Cavalleri}},\ }\bibfield  {title} {\bibinfo {title} {Polarizing an antiferromagnet by optical engineering of the crystal field},\ }\href@noop {} {\bibfield  {journal} {\bibinfo  {journal} {Nature Physics}\ }\textbf {\bibinfo {volume} {16}},\ \bibinfo {pages} {937} (\bibinfo {year} {2020})}\BibitemShut {NoStop}%
\bibitem [{\citenamefont {Shin}\ \emph {et~al.}(2018)\citenamefont {Shin}, \citenamefont {H{\"u}bener}, \citenamefont {De~Giovannini}, \citenamefont {Jin}, \citenamefont {Rubio},\ and\ \citenamefont {Park}}]{shin2018phonon}%
  \BibitemOpen
  \bibfield  {author} {\bibinfo {author} {\bibfnamefont {D.}~\bibnamefont {Shin}}, \bibinfo {author} {\bibfnamefont {H.}~\bibnamefont {H{\"u}bener}}, \bibinfo {author} {\bibfnamefont {U.}~\bibnamefont {De~Giovannini}}, \bibinfo {author} {\bibfnamefont {H.}~\bibnamefont {Jin}}, \bibinfo {author} {\bibfnamefont {A.}~\bibnamefont {Rubio}},\ and\ \bibinfo {author} {\bibfnamefont {N.}~\bibnamefont {Park}},\ }\bibfield  {title} {\bibinfo {title} {{Phonon-driven spin-Floquet magneto-valleytronics in {Mo}{S}$_2$}},\ }\href@noop {} {\bibfield  {journal} {\bibinfo  {journal} {Nature Communications}\ }\textbf {\bibinfo {volume} {9}},\ \bibinfo {pages} {638} (\bibinfo {year} {2018})}\BibitemShut {NoStop}%
\bibitem [{\citenamefont {Kogar}\ \emph {et~al.}(2020)\citenamefont {Kogar}, \citenamefont {Zong}, \citenamefont {Dolgirev}, \citenamefont {Shen}, \citenamefont {Straquadine}, \citenamefont {Bie}, \citenamefont {Wang}, \citenamefont {Rohwer}, \citenamefont {Tung}, \citenamefont {Yang}, \citenamefont {Li}, \citenamefont {Yang}, \citenamefont {Weathersby}, \citenamefont {Park}, \citenamefont {Kozina}, \citenamefont {Sie}, \citenamefont {Wen}, \citenamefont {Jarillo-Herrero}, \citenamefont {Fisher}, \citenamefont {Wang},\ and\ \citenamefont {Gedik}}]{Kogar2020light}%
  \BibitemOpen
  \bibfield  {author} {\bibinfo {author} {\bibfnamefont {A.}~\bibnamefont {Kogar}}, \bibinfo {author} {\bibfnamefont {A.}~\bibnamefont {Zong}}, \bibinfo {author} {\bibfnamefont {P.~E.}\ \bibnamefont {Dolgirev}}, \bibinfo {author} {\bibfnamefont {X.}~\bibnamefont {Shen}}, \bibinfo {author} {\bibfnamefont {J.}~\bibnamefont {Straquadine}}, \bibinfo {author} {\bibfnamefont {Y.-Q.}\ \bibnamefont {Bie}}, \bibinfo {author} {\bibfnamefont {X.}~\bibnamefont {Wang}}, \bibinfo {author} {\bibfnamefont {T.}~\bibnamefont {Rohwer}}, \bibinfo {author} {\bibfnamefont {I.-C.}\ \bibnamefont {Tung}}, \bibinfo {author} {\bibfnamefont {Y.}~\bibnamefont {Yang}}, \bibinfo {author} {\bibfnamefont {R.}~\bibnamefont {Li}}, \bibinfo {author} {\bibfnamefont {J.}~\bibnamefont {Yang}}, \bibinfo {author} {\bibfnamefont {S.}~\bibnamefont {Weathersby}}, \bibinfo {author} {\bibfnamefont {S.}~\bibnamefont {Park}}, \bibinfo {author} {\bibfnamefont {M.~E.}\ \bibnamefont {Kozina}}, \bibinfo {author} {\bibfnamefont {E.~J.}\ \bibnamefont {Sie}},
  \bibinfo {author} {\bibfnamefont {H.}~\bibnamefont {Wen}}, \bibinfo {author} {\bibfnamefont {P.}~\bibnamefont {Jarillo-Herrero}}, \bibinfo {author} {\bibfnamefont {I.~R.}\ \bibnamefont {Fisher}}, \bibinfo {author} {\bibfnamefont {X.}~\bibnamefont {Wang}},\ and\ \bibinfo {author} {\bibfnamefont {N.}~\bibnamefont {Gedik}},\ }\bibfield  {title} {\bibinfo {title} {Light-induced charge density wave in {La}{Te}$_3$},\ }\href {https://doi.org/10.1038/s41567-019-0705-3} {\bibfield  {journal} {\bibinfo  {journal} {Nature Physics}\ }\textbf {\bibinfo {volume} {16}},\ \bibinfo {pages} {159} (\bibinfo {year} {2020})}\BibitemShut {NoStop}%
\bibitem [{\citenamefont {Wang}\ \emph {et~al.}(2013)\citenamefont {Wang}, \citenamefont {Steinberg}, \citenamefont {Jarillo-Herrero},\ and\ \citenamefont {Gedik}}]{wang2013observation}%
  \BibitemOpen
  \bibfield  {author} {\bibinfo {author} {\bibfnamefont {Y.~H.}\ \bibnamefont {Wang}}, \bibinfo {author} {\bibfnamefont {H.}~\bibnamefont {Steinberg}}, \bibinfo {author} {\bibfnamefont {P.}~\bibnamefont {Jarillo-Herrero}},\ and\ \bibinfo {author} {\bibfnamefont {N.}~\bibnamefont {Gedik}},\ }\bibfield  {title} {\bibinfo {title} {{Observation of Floquet-Bloch States on the Surface of a Topological Insulator}},\ }\href {https://doi.org/10.1126/science.1239834} {\bibfield  {journal} {\bibinfo  {journal} {Science}\ }\textbf {\bibinfo {volume} {342}},\ \bibinfo {pages} {453} (\bibinfo {year} {2013})}\BibitemShut {NoStop}%
\bibitem [{\citenamefont {McIver}\ \emph {et~al.}(2020)\citenamefont {McIver}, \citenamefont {Schulte}, \citenamefont {Stein}, \citenamefont {Matsuyama}, \citenamefont {Jotzu}, \citenamefont {Meier},\ and\ \citenamefont {Cavalleri}}]{mciver2020light}%
  \BibitemOpen
  \bibfield  {author} {\bibinfo {author} {\bibfnamefont {J.~W.}\ \bibnamefont {McIver}}, \bibinfo {author} {\bibfnamefont {B.}~\bibnamefont {Schulte}}, \bibinfo {author} {\bibfnamefont {F.-U.}\ \bibnamefont {Stein}}, \bibinfo {author} {\bibfnamefont {T.}~\bibnamefont {Matsuyama}}, \bibinfo {author} {\bibfnamefont {G.}~\bibnamefont {Jotzu}}, \bibinfo {author} {\bibfnamefont {G.}~\bibnamefont {Meier}},\ and\ \bibinfo {author} {\bibfnamefont {A.}~\bibnamefont {Cavalleri}},\ }\bibfield  {title} {\bibinfo {title} {{Light-induced anomalous Hall effect in graphene}},\ }\href@noop {} {\bibfield  {journal} {\bibinfo  {journal} {Nature Physics}\ }\textbf {\bibinfo {volume} {16}},\ \bibinfo {pages} {38} (\bibinfo {year} {2020})}\BibitemShut {NoStop}%
\bibitem [{\citenamefont {Ito}\ \emph {et~al.}(2023)\citenamefont {Ito}, \citenamefont {Sch{\"u}ler}, \citenamefont {Meierhofer}, \citenamefont {Schlauderer}, \citenamefont {Freudenstein}, \citenamefont {Reimann}, \citenamefont {Afanasiev}, \citenamefont {Kokh}, \citenamefont {Tereshchenko}, \citenamefont {G{\"u}dde}, \citenamefont {Sentef}, \citenamefont {H{\"o}fer},\ and\ \citenamefont {Huber}}]{Ito2023buildup}%
  \BibitemOpen
  \bibfield  {author} {\bibinfo {author} {\bibfnamefont {S.}~\bibnamefont {Ito}}, \bibinfo {author} {\bibfnamefont {M.}~\bibnamefont {Sch{\"u}ler}}, \bibinfo {author} {\bibfnamefont {M.}~\bibnamefont {Meierhofer}}, \bibinfo {author} {\bibfnamefont {S.}~\bibnamefont {Schlauderer}}, \bibinfo {author} {\bibfnamefont {J.}~\bibnamefont {Freudenstein}}, \bibinfo {author} {\bibfnamefont {J.}~\bibnamefont {Reimann}}, \bibinfo {author} {\bibfnamefont {D.}~\bibnamefont {Afanasiev}}, \bibinfo {author} {\bibfnamefont {K.~A.}\ \bibnamefont {Kokh}}, \bibinfo {author} {\bibfnamefont {O.~E.}\ \bibnamefont {Tereshchenko}}, \bibinfo {author} {\bibfnamefont {J.}~\bibnamefont {G{\"u}dde}}, \bibinfo {author} {\bibfnamefont {M.~A.}\ \bibnamefont {Sentef}}, \bibinfo {author} {\bibfnamefont {U.}~\bibnamefont {H{\"o}fer}},\ and\ \bibinfo {author} {\bibfnamefont {R.}~\bibnamefont {Huber}},\ }\bibfield  {title} {\bibinfo {title} {{Build-up and dephasing of Floquet--Bloch bands on subcycle timescales}},\ }\href
  {https://doi.org/10.1038/s41586-023-05850-x} {\bibfield  {journal} {\bibinfo  {journal} {Nature}\ }\textbf {\bibinfo {volume} {616}},\ \bibinfo {pages} {696} (\bibinfo {year} {2023})}\BibitemShut {NoStop}%
\bibitem [{\citenamefont {Fausti}\ \emph {et~al.}(2011)\citenamefont {Fausti}, \citenamefont {Tobey}, \citenamefont {Dean}, \citenamefont {Kaiser}, \citenamefont {Dienst}, \citenamefont {Hoffmann}, \citenamefont {Pyon}, \citenamefont {Takayama}, \citenamefont {Takagi},\ and\ \citenamefont {Cavalleri}}]{fausti2011light}%
  \BibitemOpen
  \bibfield  {author} {\bibinfo {author} {\bibfnamefont {D.}~\bibnamefont {Fausti}}, \bibinfo {author} {\bibfnamefont {R.}~\bibnamefont {Tobey}}, \bibinfo {author} {\bibfnamefont {N.}~\bibnamefont {Dean}}, \bibinfo {author} {\bibfnamefont {S.}~\bibnamefont {Kaiser}}, \bibinfo {author} {\bibfnamefont {A.}~\bibnamefont {Dienst}}, \bibinfo {author} {\bibfnamefont {M.~C.}\ \bibnamefont {Hoffmann}}, \bibinfo {author} {\bibfnamefont {S.}~\bibnamefont {Pyon}}, \bibinfo {author} {\bibfnamefont {T.}~\bibnamefont {Takayama}}, \bibinfo {author} {\bibfnamefont {H.}~\bibnamefont {Takagi}},\ and\ \bibinfo {author} {\bibfnamefont {A.}~\bibnamefont {Cavalleri}},\ }\bibfield  {title} {\bibinfo {title} {Light-induced superconductivity in a stripe-ordered cuprate},\ }\href@noop {} {\bibfield  {journal} {\bibinfo  {journal} {Science}\ }\textbf {\bibinfo {volume} {331}},\ \bibinfo {pages} {189} (\bibinfo {year} {2011})}\BibitemShut {NoStop}%
\bibitem [{\citenamefont {Hu}\ \emph {et~al.}(2014)\citenamefont {Hu}, \citenamefont {Kaiser}, \citenamefont {Nicoletti}, \citenamefont {Hunt}, \citenamefont {Gierz}, \citenamefont {Hoffmann}, \citenamefont {Le~Tacon}, \citenamefont {Loew}, \citenamefont {Keimer},\ and\ \citenamefont {Cavalleri}}]{hu2014optically}%
  \BibitemOpen
  \bibfield  {author} {\bibinfo {author} {\bibfnamefont {W.}~\bibnamefont {Hu}}, \bibinfo {author} {\bibfnamefont {S.}~\bibnamefont {Kaiser}}, \bibinfo {author} {\bibfnamefont {D.}~\bibnamefont {Nicoletti}}, \bibinfo {author} {\bibfnamefont {C.~R.}\ \bibnamefont {Hunt}}, \bibinfo {author} {\bibfnamefont {I.}~\bibnamefont {Gierz}}, \bibinfo {author} {\bibfnamefont {M.~C.}\ \bibnamefont {Hoffmann}}, \bibinfo {author} {\bibfnamefont {M.}~\bibnamefont {Le~Tacon}}, \bibinfo {author} {\bibfnamefont {T.}~\bibnamefont {Loew}}, \bibinfo {author} {\bibfnamefont {B.}~\bibnamefont {Keimer}},\ and\ \bibinfo {author} {\bibfnamefont {A.}~\bibnamefont {Cavalleri}},\ }\bibfield  {title} {\bibinfo {title} {Optically enhanced coherent transport in {Y}{Ba}$_2${Cu}$_3${O}$_{6.5}$ by ultrafast redistribution of interlayer coupling},\ }\href@noop {} {\bibfield  {journal} {\bibinfo  {journal} {Nature Materials}\ }\textbf {\bibinfo {volume} {13}},\ \bibinfo {pages} {705} (\bibinfo {year} {2014})}\BibitemShut {NoStop}%
\bibitem [{\citenamefont {Mitrano}\ \emph {et~al.}(2016)\citenamefont {Mitrano}, \citenamefont {Cantaluppi}, \citenamefont {Nicoletti}, \citenamefont {Kaiser}, \citenamefont {Perucchi}, \citenamefont {Lupi}, \citenamefont {Di~Pietro}, \citenamefont {Pontiroli}, \citenamefont {Ricc{\`o}}, \citenamefont {Clark} \emph {et~al.}}]{mitrano2016possible}%
  \BibitemOpen
  \bibfield  {author} {\bibinfo {author} {\bibfnamefont {M.}~\bibnamefont {Mitrano}}, \bibinfo {author} {\bibfnamefont {A.}~\bibnamefont {Cantaluppi}}, \bibinfo {author} {\bibfnamefont {D.}~\bibnamefont {Nicoletti}}, \bibinfo {author} {\bibfnamefont {S.}~\bibnamefont {Kaiser}}, \bibinfo {author} {\bibfnamefont {A.}~\bibnamefont {Perucchi}}, \bibinfo {author} {\bibfnamefont {S.}~\bibnamefont {Lupi}}, \bibinfo {author} {\bibfnamefont {P.}~\bibnamefont {Di~Pietro}}, \bibinfo {author} {\bibfnamefont {D.}~\bibnamefont {Pontiroli}}, \bibinfo {author} {\bibfnamefont {M.}~\bibnamefont {Ricc{\`o}}}, \bibinfo {author} {\bibfnamefont {S.~R.}\ \bibnamefont {Clark}}, \emph {et~al.},\ }\bibfield  {title} {\bibinfo {title} {{Possible light-induced superconductivity in {K}$_3${C}$_{60}$ at high temperature}},\ }\href@noop {} {\bibfield  {journal} {\bibinfo  {journal} {Nature}\ }\textbf {\bibinfo {volume} {530}},\ \bibinfo {pages} {461} (\bibinfo {year} {2016})}\BibitemShut {NoStop}%
\bibitem [{\citenamefont {von~der Linde}\ \emph {et~al.}(1974)\citenamefont {von~der Linde}, \citenamefont {Glass},\ and\ \citenamefont {Rodgers}}]{vonderlinde1974multiphoton}%
  \BibitemOpen
  \bibfield  {author} {\bibinfo {author} {\bibfnamefont {D.}~\bibnamefont {von~der Linde}}, \bibinfo {author} {\bibfnamefont {A.~M.}\ \bibnamefont {Glass}},\ and\ \bibinfo {author} {\bibfnamefont {K.~F.}\ \bibnamefont {Rodgers}},\ }\bibfield  {title} {\bibinfo {title} {{Multiphoton photorefractive processes for optical storage in LiNbO$_3$}},\ }\href {https://doi.org/10.1063/1.1655420} {\bibfield  {journal} {\bibinfo  {journal} {Applied Physics Letters}\ }\textbf {\bibinfo {volume} {25}},\ \bibinfo {pages} {155} (\bibinfo {year} {1974})}\BibitemShut {NoStop}%
\bibitem [{\citenamefont {Koshihara}\ \emph {et~al.}(1990)\citenamefont {Koshihara}, \citenamefont {Tokura}, \citenamefont {Mitani}, \citenamefont {Saito},\ and\ \citenamefont {Koda}}]{Koshihara1990photoinduced}%
  \BibitemOpen
  \bibfield  {author} {\bibinfo {author} {\bibfnamefont {S.}~\bibnamefont {Koshihara}}, \bibinfo {author} {\bibfnamefont {Y.}~\bibnamefont {Tokura}}, \bibinfo {author} {\bibfnamefont {T.}~\bibnamefont {Mitani}}, \bibinfo {author} {\bibfnamefont {G.}~\bibnamefont {Saito}},\ and\ \bibinfo {author} {\bibfnamefont {T.}~\bibnamefont {Koda}},\ }\bibfield  {title} {\bibinfo {title} {{Photoinduced valence instability in the organic molecular compound tetrathiafulvalene-p-chloranil (TTF-CA)}},\ }\href {https://doi.org/10.1103/PhysRevB.42.6853} {\bibfield  {journal} {\bibinfo  {journal} {Physical Review B}\ }\textbf {\bibinfo {volume} {42}},\ \bibinfo {pages} {6853} (\bibinfo {year} {1990})}\BibitemShut {NoStop}%
\bibitem [{\citenamefont {Kiryukhin}\ \emph {et~al.}(1997)\citenamefont {Kiryukhin}, \citenamefont {Casa}, \citenamefont {Hill}, \citenamefont {Keimer}, \citenamefont {Vigliante}, \citenamefont {Tomioka},\ and\ \citenamefont {Tokura}}]{Kiryukhin1997xray}%
  \BibitemOpen
  \bibfield  {author} {\bibinfo {author} {\bibfnamefont {V.}~\bibnamefont {Kiryukhin}}, \bibinfo {author} {\bibfnamefont {D.}~\bibnamefont {Casa}}, \bibinfo {author} {\bibfnamefont {J.~P.}\ \bibnamefont {Hill}}, \bibinfo {author} {\bibfnamefont {B.}~\bibnamefont {Keimer}}, \bibinfo {author} {\bibfnamefont {A.}~\bibnamefont {Vigliante}}, \bibinfo {author} {\bibfnamefont {Y.}~\bibnamefont {Tomioka}},\ and\ \bibinfo {author} {\bibfnamefont {Y.}~\bibnamefont {Tokura}},\ }\bibfield  {title} {\bibinfo {title} {An x-ray-induced insulator--metal transition in a magnetoresistive manganite},\ }\href {https://doi.org/10.1038/386813a0} {\bibfield  {journal} {\bibinfo  {journal} {Nature}\ }\textbf {\bibinfo {volume} {386}},\ \bibinfo {pages} {813} (\bibinfo {year} {1997})}\BibitemShut {NoStop}%
\bibitem [{\citenamefont {Fiebig}\ \emph {et~al.}(1998)\citenamefont {Fiebig}, \citenamefont {Miyano}, \citenamefont {Tomioka},\ and\ \citenamefont {Tokura}}]{fiebig1998visualization}%
  \BibitemOpen
  \bibfield  {author} {\bibinfo {author} {\bibfnamefont {M.}~\bibnamefont {Fiebig}}, \bibinfo {author} {\bibfnamefont {K.}~\bibnamefont {Miyano}}, \bibinfo {author} {\bibfnamefont {Y.}~\bibnamefont {Tomioka}},\ and\ \bibinfo {author} {\bibfnamefont {Y.}~\bibnamefont {Tokura}},\ }\bibfield  {title} {\bibinfo {title} {Visualization of the local insulator-metal transition in {Pr}$_{0.7}${Ca}$_{0.3}${Mn}{O}$_3$},\ }\href@noop {} {\bibfield  {journal} {\bibinfo  {journal} {Science}\ }\textbf {\bibinfo {volume} {280}},\ \bibinfo {pages} {1925} (\bibinfo {year} {1998})}\BibitemShut {NoStop}%
\bibitem [{\citenamefont {Rini}\ \emph {et~al.}(2007)\citenamefont {Rini}, \citenamefont {Tobey}, \citenamefont {Dean}, \citenamefont {Itatani}, \citenamefont {Tomioka}, \citenamefont {Tokura}, \citenamefont {Schoenlein},\ and\ \citenamefont {Cavalleri}}]{rini2007control}%
  \BibitemOpen
  \bibfield  {author} {\bibinfo {author} {\bibfnamefont {M.}~\bibnamefont {Rini}}, \bibinfo {author} {\bibfnamefont {R.}~\bibnamefont {Tobey}}, \bibinfo {author} {\bibfnamefont {N.}~\bibnamefont {Dean}}, \bibinfo {author} {\bibfnamefont {J.}~\bibnamefont {Itatani}}, \bibinfo {author} {\bibfnamefont {Y.}~\bibnamefont {Tomioka}}, \bibinfo {author} {\bibfnamefont {Y.}~\bibnamefont {Tokura}}, \bibinfo {author} {\bibfnamefont {R.~W.}\ \bibnamefont {Schoenlein}},\ and\ \bibinfo {author} {\bibfnamefont {A.}~\bibnamefont {Cavalleri}},\ }\bibfield  {title} {\bibinfo {title} {Control of the electronic phase of a manganite by mode-selective vibrational excitation},\ }\href@noop {} {\bibfield  {journal} {\bibinfo  {journal} {Nature}\ }\textbf {\bibinfo {volume} {449}},\ \bibinfo {pages} {72} (\bibinfo {year} {2007})}\BibitemShut {NoStop}%
\bibitem [{\citenamefont {Ichikawa}\ \emph {et~al.}(2011)\citenamefont {Ichikawa}, \citenamefont {Nozawa}, \citenamefont {Sato}, \citenamefont {Tomita}, \citenamefont {Ichiyanagi}, \citenamefont {Chollet}, \citenamefont {Guerin}, \citenamefont {Dean}, \citenamefont {Cavalleri}, \citenamefont {Adachi}, \citenamefont {Arima}, \citenamefont {Sawa}, \citenamefont {Ogimoto}, \citenamefont {Nakamura}, \citenamefont {Tamaki}, \citenamefont {Miyano},\ and\ \citenamefont {Koshihara}}]{Ichikawa2011transient}%
  \BibitemOpen
  \bibfield  {author} {\bibinfo {author} {\bibfnamefont {H.}~\bibnamefont {Ichikawa}}, \bibinfo {author} {\bibfnamefont {S.}~\bibnamefont {Nozawa}}, \bibinfo {author} {\bibfnamefont {T.}~\bibnamefont {Sato}}, \bibinfo {author} {\bibfnamefont {A.}~\bibnamefont {Tomita}}, \bibinfo {author} {\bibfnamefont {K.}~\bibnamefont {Ichiyanagi}}, \bibinfo {author} {\bibfnamefont {M.}~\bibnamefont {Chollet}}, \bibinfo {author} {\bibfnamefont {L.}~\bibnamefont {Guerin}}, \bibinfo {author} {\bibfnamefont {N.}~\bibnamefont {Dean}}, \bibinfo {author} {\bibfnamefont {A.}~\bibnamefont {Cavalleri}}, \bibinfo {author} {\bibfnamefont {S.-i.}\ \bibnamefont {Adachi}}, \bibinfo {author} {\bibfnamefont {T.-h.}\ \bibnamefont {Arima}}, \bibinfo {author} {\bibfnamefont {H.}~\bibnamefont {Sawa}}, \bibinfo {author} {\bibfnamefont {Y.}~\bibnamefont {Ogimoto}}, \bibinfo {author} {\bibfnamefont {M.}~\bibnamefont {Nakamura}}, \bibinfo {author} {\bibfnamefont {R.}~\bibnamefont {Tamaki}}, \bibinfo {author} {\bibfnamefont {K.}~\bibnamefont
  {Miyano}},\ and\ \bibinfo {author} {\bibfnamefont {S.-y.}\ \bibnamefont {Koshihara}},\ }\bibfield  {title} {\bibinfo {title} {{Transient photoinduced `hidden' phase in a manganite}},\ }\href {https://doi.org/10.1038/nmat2929} {\bibfield  {journal} {\bibinfo  {journal} {Nature Materials}\ }\textbf {\bibinfo {volume} {10}},\ \bibinfo {pages} {101} (\bibinfo {year} {2011})}\BibitemShut {NoStop}%
\bibitem [{\citenamefont {Zhang}\ \emph {et~al.}(2016)\citenamefont {Zhang}, \citenamefont {Tan}, \citenamefont {Liu}, \citenamefont {Teitelbaum}, \citenamefont {Post}, \citenamefont {Jin}, \citenamefont {Nelson}, \citenamefont {Basov}, \citenamefont {Wu},\ and\ \citenamefont {Averitt}}]{zhang2016cooperative}%
  \BibitemOpen
  \bibfield  {author} {\bibinfo {author} {\bibfnamefont {J.}~\bibnamefont {Zhang}}, \bibinfo {author} {\bibfnamefont {X.}~\bibnamefont {Tan}}, \bibinfo {author} {\bibfnamefont {M.}~\bibnamefont {Liu}}, \bibinfo {author} {\bibfnamefont {S.~W.}\ \bibnamefont {Teitelbaum}}, \bibinfo {author} {\bibfnamefont {K.~W.}\ \bibnamefont {Post}}, \bibinfo {author} {\bibfnamefont {F.}~\bibnamefont {Jin}}, \bibinfo {author} {\bibfnamefont {K.~A.}\ \bibnamefont {Nelson}}, \bibinfo {author} {\bibfnamefont {D.~N.}\ \bibnamefont {Basov}}, \bibinfo {author} {\bibfnamefont {W.}~\bibnamefont {Wu}},\ and\ \bibinfo {author} {\bibfnamefont {R.~D.}\ \bibnamefont {Averitt}},\ }\bibfield  {title} {\bibinfo {title} {Cooperative photoinduced metastable phase control in strained manganite films},\ }\href@noop {} {\bibfield  {journal} {\bibinfo  {journal} {Nature Materials}\ }\textbf {\bibinfo {volume} {15}},\ \bibinfo {pages} {956} (\bibinfo {year} {2016})}\BibitemShut {NoStop}%
\bibitem [{\citenamefont {Stojchevska}\ \emph {et~al.}(2014)\citenamefont {Stojchevska}, \citenamefont {Vaskivskyi}, \citenamefont {Mertelj}, \citenamefont {Kusar}, \citenamefont {Svetin}, \citenamefont {Brazovskii},\ and\ \citenamefont {Mihailovic}}]{Stojchevska2015ultrafast}%
  \BibitemOpen
  \bibfield  {author} {\bibinfo {author} {\bibfnamefont {L.}~\bibnamefont {Stojchevska}}, \bibinfo {author} {\bibfnamefont {I.}~\bibnamefont {Vaskivskyi}}, \bibinfo {author} {\bibfnamefont {T.}~\bibnamefont {Mertelj}}, \bibinfo {author} {\bibfnamefont {P.}~\bibnamefont {Kusar}}, \bibinfo {author} {\bibfnamefont {D.}~\bibnamefont {Svetin}}, \bibinfo {author} {\bibfnamefont {S.}~\bibnamefont {Brazovskii}},\ and\ \bibinfo {author} {\bibfnamefont {D.}~\bibnamefont {Mihailovic}},\ }\bibfield  {title} {\bibinfo {title} {Ultrafast switching to a stable hidden quantum state in an electronic crystal},\ }\href {https://doi.org/10.1126/science.1241591} {\bibfield  {journal} {\bibinfo  {journal} {Science}\ }\textbf {\bibinfo {volume} {344}},\ \bibinfo {pages} {177} (\bibinfo {year} {2014})}\BibitemShut {NoStop}%
\bibitem [{\citenamefont {Stoica}\ \emph {et~al.}(2019)\citenamefont {Stoica}, \citenamefont {Laanait}, \citenamefont {Dai}, \citenamefont {Hong}, \citenamefont {Yuan}, \citenamefont {Zhang}, \citenamefont {Lei}, \citenamefont {McCarter}, \citenamefont {Yadav}, \citenamefont {Damodaran} \emph {et~al.}}]{stoica2019optical}%
  \BibitemOpen
  \bibfield  {author} {\bibinfo {author} {\bibfnamefont {V.}~\bibnamefont {Stoica}}, \bibinfo {author} {\bibfnamefont {N.}~\bibnamefont {Laanait}}, \bibinfo {author} {\bibfnamefont {C.}~\bibnamefont {Dai}}, \bibinfo {author} {\bibfnamefont {Z.}~\bibnamefont {Hong}}, \bibinfo {author} {\bibfnamefont {Y.}~\bibnamefont {Yuan}}, \bibinfo {author} {\bibfnamefont {Z.}~\bibnamefont {Zhang}}, \bibinfo {author} {\bibfnamefont {S.}~\bibnamefont {Lei}}, \bibinfo {author} {\bibfnamefont {M.}~\bibnamefont {McCarter}}, \bibinfo {author} {\bibfnamefont {A.}~\bibnamefont {Yadav}}, \bibinfo {author} {\bibfnamefont {A.}~\bibnamefont {Damodaran}}, \emph {et~al.},\ }\bibfield  {title} {\bibinfo {title} {Optical creation of a supercrystal with three-dimensional nanoscale periodicity},\ }\href@noop {} {\bibfield  {journal} {\bibinfo  {journal} {Nature Materials}\ }\textbf {\bibinfo {volume} {18}},\ \bibinfo {pages} {377} (\bibinfo {year} {2019})}\BibitemShut {NoStop}%
\bibitem [{\citenamefont {Nova}\ \emph {et~al.}(2019)\citenamefont {Nova}, \citenamefont {Disa}, \citenamefont {Fechner},\ and\ \citenamefont {Cavalleri}}]{Nova2019metastable}%
  \BibitemOpen
  \bibfield  {author} {\bibinfo {author} {\bibfnamefont {T.~F.}\ \bibnamefont {Nova}}, \bibinfo {author} {\bibfnamefont {A.~S.}\ \bibnamefont {Disa}}, \bibinfo {author} {\bibfnamefont {M.}~\bibnamefont {Fechner}},\ and\ \bibinfo {author} {\bibfnamefont {A.}~\bibnamefont {Cavalleri}},\ }\bibfield  {title} {\bibinfo {title} {Metastable ferroelectricity in optically strained {SrTiO$_3$}},\ }\href {https://doi.org/10.1126/science.aaw4911} {\bibfield  {journal} {\bibinfo  {journal} {Science}\ }\textbf {\bibinfo {volume} {364}},\ \bibinfo {pages} {1075} (\bibinfo {year} {2019})}\BibitemShut {NoStop}%
\bibitem [{\citenamefont {Disa}\ \emph {et~al.}(2023)\citenamefont {Disa}, \citenamefont {Curtis}, \citenamefont {Fechner}, \citenamefont {Liu}, \citenamefont {Von~Hoegen}, \citenamefont {F{\"o}rst}, \citenamefont {Nova}, \citenamefont {Narang}, \citenamefont {Maljuk}, \citenamefont {Boris} \emph {et~al.}}]{disa2023photo}%
  \BibitemOpen
  \bibfield  {author} {\bibinfo {author} {\bibfnamefont {A.}~\bibnamefont {Disa}}, \bibinfo {author} {\bibfnamefont {J.}~\bibnamefont {Curtis}}, \bibinfo {author} {\bibfnamefont {M.}~\bibnamefont {Fechner}}, \bibinfo {author} {\bibfnamefont {A.}~\bibnamefont {Liu}}, \bibinfo {author} {\bibfnamefont {A.}~\bibnamefont {Von~Hoegen}}, \bibinfo {author} {\bibfnamefont {M.}~\bibnamefont {F{\"o}rst}}, \bibinfo {author} {\bibfnamefont {T.}~\bibnamefont {Nova}}, \bibinfo {author} {\bibfnamefont {P.}~\bibnamefont {Narang}}, \bibinfo {author} {\bibfnamefont {A.}~\bibnamefont {Maljuk}}, \bibinfo {author} {\bibfnamefont {A.}~\bibnamefont {Boris}}, \emph {et~al.},\ }\bibfield  {title} {\bibinfo {title} {Photo-induced high-temperature ferromagnetism in {Y}{Ti}{O}$_3$},\ }\href@noop {} {\bibfield  {journal} {\bibinfo  {journal} {Nature}\ }\textbf {\bibinfo {volume} {617}},\ \bibinfo {pages} {73} (\bibinfo {year} {2023})}\BibitemShut {NoStop}%
\bibitem [{\citenamefont {Vogelgesang}\ \emph {et~al.}(2018)\citenamefont {Vogelgesang}, \citenamefont {Storeck}, \citenamefont {Horstmann}, \citenamefont {Diekmann}, \citenamefont {Sivis}, \citenamefont {Schramm}, \citenamefont {Rossnagel}, \citenamefont {Sch{\"a}fer},\ and\ \citenamefont {Ropers}}]{Vogelgesang2018phase}%
  \BibitemOpen
  \bibfield  {author} {\bibinfo {author} {\bibfnamefont {S.}~\bibnamefont {Vogelgesang}}, \bibinfo {author} {\bibfnamefont {G.}~\bibnamefont {Storeck}}, \bibinfo {author} {\bibfnamefont {J.~G.}\ \bibnamefont {Horstmann}}, \bibinfo {author} {\bibfnamefont {T.}~\bibnamefont {Diekmann}}, \bibinfo {author} {\bibfnamefont {M.}~\bibnamefont {Sivis}}, \bibinfo {author} {\bibfnamefont {S.}~\bibnamefont {Schramm}}, \bibinfo {author} {\bibfnamefont {K.}~\bibnamefont {Rossnagel}}, \bibinfo {author} {\bibfnamefont {S.}~\bibnamefont {Sch{\"a}fer}},\ and\ \bibinfo {author} {\bibfnamefont {C.}~\bibnamefont {Ropers}},\ }\bibfield  {title} {\bibinfo {title} {Phase ordering of charge density waves traced by ultrafast low-energy electron diffraction},\ }\href {https://doi.org/10.1038/nphys4309} {\bibfield  {journal} {\bibinfo  {journal} {Nature Physics}\ }\textbf {\bibinfo {volume} {14}},\ \bibinfo {pages} {184} (\bibinfo {year} {2018})}\BibitemShut {NoStop}%
\bibitem [{\citenamefont {Zong}\ \emph {et~al.}(2019)\citenamefont {Zong}, \citenamefont {Kogar}, \citenamefont {Bie}, \citenamefont {Rohwer}, \citenamefont {Lee}, \citenamefont {Baldini}, \citenamefont {Erge{\c{c}}en}, \citenamefont {Yilmaz}, \citenamefont {Freelon}, \citenamefont {Sie} \emph {et~al.}}]{zong2019evidence}%
  \BibitemOpen
  \bibfield  {author} {\bibinfo {author} {\bibfnamefont {A.}~\bibnamefont {Zong}}, \bibinfo {author} {\bibfnamefont {A.}~\bibnamefont {Kogar}}, \bibinfo {author} {\bibfnamefont {Y.-Q.}\ \bibnamefont {Bie}}, \bibinfo {author} {\bibfnamefont {T.}~\bibnamefont {Rohwer}}, \bibinfo {author} {\bibfnamefont {C.}~\bibnamefont {Lee}}, \bibinfo {author} {\bibfnamefont {E.}~\bibnamefont {Baldini}}, \bibinfo {author} {\bibfnamefont {E.}~\bibnamefont {Erge{\c{c}}en}}, \bibinfo {author} {\bibfnamefont {M.~B.}\ \bibnamefont {Yilmaz}}, \bibinfo {author} {\bibfnamefont {B.}~\bibnamefont {Freelon}}, \bibinfo {author} {\bibfnamefont {E.~J.}\ \bibnamefont {Sie}}, \emph {et~al.},\ }\bibfield  {title} {\bibinfo {title} {Evidence for topological defects in a photoinduced phase transition},\ }\href@noop {} {\bibfield  {journal} {\bibinfo  {journal} {Nature Physics}\ }\textbf {\bibinfo {volume} {15}},\ \bibinfo {pages} {27} (\bibinfo {year} {2019})}\BibitemShut {NoStop}%
\bibitem [{\citenamefont {Gerasimenko}\ \emph {et~al.}(2019)\citenamefont {Gerasimenko}, \citenamefont {Vaskivskyi}, \citenamefont {Litskevich}, \citenamefont {Ravnik}, \citenamefont {Vodeb}, \citenamefont {Diego}, \citenamefont {Kabanov},\ and\ \citenamefont {Mihailovic}}]{Gerasimenko2019quantum}%
  \BibitemOpen
  \bibfield  {author} {\bibinfo {author} {\bibfnamefont {Y.~A.}\ \bibnamefont {Gerasimenko}}, \bibinfo {author} {\bibfnamefont {I.}~\bibnamefont {Vaskivskyi}}, \bibinfo {author} {\bibfnamefont {M.}~\bibnamefont {Litskevich}}, \bibinfo {author} {\bibfnamefont {J.}~\bibnamefont {Ravnik}}, \bibinfo {author} {\bibfnamefont {J.}~\bibnamefont {Vodeb}}, \bibinfo {author} {\bibfnamefont {M.}~\bibnamefont {Diego}}, \bibinfo {author} {\bibfnamefont {V.}~\bibnamefont {Kabanov}},\ and\ \bibinfo {author} {\bibfnamefont {D.}~\bibnamefont {Mihailovic}},\ }\bibfield  {title} {\bibinfo {title} {Quantum jamming transition to a correlated electron glass in {1T}-{Ta}{S}$_2$},\ }\href {https://doi.org/10.1038/s41563-019-0423-3} {\bibfield  {journal} {\bibinfo  {journal} {Nature Materials}\ }\textbf {\bibinfo {volume} {18}},\ \bibinfo {pages} {1078} (\bibinfo {year} {2019})}\BibitemShut {NoStop}%
\bibitem [{\citenamefont {Budden}\ \emph {et~al.}(2021)\citenamefont {Budden}, \citenamefont {Gebert}, \citenamefont {Buzzi}, \citenamefont {Jotzu}, \citenamefont {Wang}, \citenamefont {Matsuyama}, \citenamefont {Meier}, \citenamefont {Laplace}, \citenamefont {Pontiroli}, \citenamefont {Ricc{\`o}} \emph {et~al.}}]{budden2021evidence}%
  \BibitemOpen
  \bibfield  {author} {\bibinfo {author} {\bibfnamefont {M.}~\bibnamefont {Budden}}, \bibinfo {author} {\bibfnamefont {T.}~\bibnamefont {Gebert}}, \bibinfo {author} {\bibfnamefont {M.}~\bibnamefont {Buzzi}}, \bibinfo {author} {\bibfnamefont {G.}~\bibnamefont {Jotzu}}, \bibinfo {author} {\bibfnamefont {E.}~\bibnamefont {Wang}}, \bibinfo {author} {\bibfnamefont {T.}~\bibnamefont {Matsuyama}}, \bibinfo {author} {\bibfnamefont {G.}~\bibnamefont {Meier}}, \bibinfo {author} {\bibfnamefont {Y.}~\bibnamefont {Laplace}}, \bibinfo {author} {\bibfnamefont {D.}~\bibnamefont {Pontiroli}}, \bibinfo {author} {\bibfnamefont {M.}~\bibnamefont {Ricc{\`o}}}, \emph {et~al.},\ }\bibfield  {title} {\bibinfo {title} {Evidence for metastable photo-induced superconductivity in {K}$_3${C}$_{60}$},\ }\href@noop {} {\bibfield  {journal} {\bibinfo  {journal} {Nature Physics}\ }\textbf {\bibinfo {volume} {17}},\ \bibinfo {pages} {611} (\bibinfo {year} {2021})}\BibitemShut {NoStop}%
\bibitem [{\citenamefont {Murakami}\ \emph {et~al.}(2025)\citenamefont {Murakami}, \citenamefont {Gole{\v{z}}}, \citenamefont {Eckstein},\ and\ \citenamefont {Werner}}]{murakami2025photoinduced}%
  \BibitemOpen
  \bibfield  {author} {\bibinfo {author} {\bibfnamefont {Y.}~\bibnamefont {Murakami}}, \bibinfo {author} {\bibfnamefont {D.}~\bibnamefont {Gole{\v{z}}}}, \bibinfo {author} {\bibfnamefont {M.}~\bibnamefont {Eckstein}},\ and\ \bibinfo {author} {\bibfnamefont {P.}~\bibnamefont {Werner}},\ }\bibfield  {title} {\bibinfo {title} {{Photoinduced nonequilibrium states in Mott insulators}},\ }\href@noop {} {\bibfield  {journal} {\bibinfo  {journal} {Reviews of Modern Physics}\ }\textbf {\bibinfo {volume} {97}},\ \bibinfo {pages} {035001} (\bibinfo {year} {2025})}\BibitemShut {NoStop}%
\bibitem [{\citenamefont {Kollath}\ \emph {et~al.}(2007)\citenamefont {Kollath}, \citenamefont {L{\"a}uchli},\ and\ \citenamefont {Altman}}]{kollath2007quench}%
  \BibitemOpen
  \bibfield  {author} {\bibinfo {author} {\bibfnamefont {C.}~\bibnamefont {Kollath}}, \bibinfo {author} {\bibfnamefont {A.~M.}\ \bibnamefont {L{\"a}uchli}},\ and\ \bibinfo {author} {\bibfnamefont {E.}~\bibnamefont {Altman}},\ }\bibfield  {title} {\bibinfo {title} {{Quench dynamics and nonequilibrium phase diagram of the Bose-Hubbard model}},\ }\href@noop {} {\bibfield  {journal} {\bibinfo  {journal} {Physical Review Letters}\ }\textbf {\bibinfo {volume} {98}},\ \bibinfo {pages} {180601} (\bibinfo {year} {2007})}\BibitemShut {NoStop}%
\bibitem [{\citenamefont {Eckstein}\ \emph {et~al.}(2009)\citenamefont {Eckstein}, \citenamefont {Kollar},\ and\ \citenamefont {Werner}}]{eckstein2009thermalization}%
  \BibitemOpen
  \bibfield  {author} {\bibinfo {author} {\bibfnamefont {M.}~\bibnamefont {Eckstein}}, \bibinfo {author} {\bibfnamefont {M.}~\bibnamefont {Kollar}},\ and\ \bibinfo {author} {\bibfnamefont {P.}~\bibnamefont {Werner}},\ }\bibfield  {title} {\bibinfo {title} {{Thermalization after an interaction quench in the Hubbard model}},\ }\href@noop {} {\bibfield  {journal} {\bibinfo  {journal} {Physical Review Letters}\ }\textbf {\bibinfo {volume} {103}},\ \bibinfo {pages} {056403} (\bibinfo {year} {2009})}\BibitemShut {NoStop}%
\bibitem [{\citenamefont {Kollar}\ \emph {et~al.}(2011)\citenamefont {Kollar}, \citenamefont {Wolf},\ and\ \citenamefont {Eckstein}}]{kollar2011generalized}%
  \BibitemOpen
  \bibfield  {author} {\bibinfo {author} {\bibfnamefont {M.}~\bibnamefont {Kollar}}, \bibinfo {author} {\bibfnamefont {F.~A.}\ \bibnamefont {Wolf}},\ and\ \bibinfo {author} {\bibfnamefont {M.}~\bibnamefont {Eckstein}},\ }\bibfield  {title} {\bibinfo {title} {Generalized gibbs ensemble prediction of prethermalization plateaus and their relation to nonthermal steady states in integrable systems},\ }\href@noop {} {\bibfield  {journal} {\bibinfo  {journal} {Physical Review B}\ }\textbf {\bibinfo {volume} {84}},\ \bibinfo {pages} {054304} (\bibinfo {year} {2011})}\BibitemShut {NoStop}%
\bibitem [{\citenamefont {Sun}\ and\ \citenamefont {Millis}(2020)}]{sun2020transient}%
  \BibitemOpen
  \bibfield  {author} {\bibinfo {author} {\bibfnamefont {Z.}~\bibnamefont {Sun}}\ and\ \bibinfo {author} {\bibfnamefont {A.~J.}\ \bibnamefont {Millis}},\ }\bibfield  {title} {\bibinfo {title} {Transient trapping into metastable states in systems with competing orders},\ }\href@noop {} {\bibfield  {journal} {\bibinfo  {journal} {Physical Review X}\ }\textbf {\bibinfo {volume} {10}},\ \bibinfo {pages} {021028} (\bibinfo {year} {2020})}\BibitemShut {NoStop}%
\bibitem [{\citenamefont {Masoumi}\ \emph {et~al.}(2025)\citenamefont {Masoumi}, \citenamefont {de~la Torre},\ and\ \citenamefont {Fiete}}]{Masoumi2025metastability}%
  \BibitemOpen
  \bibfield  {author} {\bibinfo {author} {\bibfnamefont {Y.}~\bibnamefont {Masoumi}}, \bibinfo {author} {\bibfnamefont {A.}~\bibnamefont {de~la Torre}},\ and\ \bibinfo {author} {\bibfnamefont {G.~A.}\ \bibnamefont {Fiete}},\ }\bibfield  {title} {\bibinfo {title} {Metastability in coexisting competing orders},\ }\href {https://doi.org/10.1103/3dm4-p2yq} {\bibfield  {journal} {\bibinfo  {journal} {Physical Review Letters}\ }\textbf {\bibinfo {volume} {135}},\ \bibinfo {pages} {066501} (\bibinfo {year} {2025})}\BibitemShut {NoStop}%
\bibitem [{\citenamefont {Oka}\ and\ \citenamefont {Kitamura}(2019)}]{oka2019floquet}%
  \BibitemOpen
  \bibfield  {author} {\bibinfo {author} {\bibfnamefont {T.}~\bibnamefont {Oka}}\ and\ \bibinfo {author} {\bibfnamefont {S.}~\bibnamefont {Kitamura}},\ }\bibfield  {title} {\bibinfo {title} {{Floquet engineering of quantum materials}},\ }\href@noop {} {\bibfield  {journal} {\bibinfo  {journal} {Annual Review of Condensed Matter Physics}\ }\textbf {\bibinfo {volume} {10}},\ \bibinfo {pages} {387} (\bibinfo {year} {2019})}\BibitemShut {NoStop}%
\bibitem [{\citenamefont {Kune{\v{s}}}(2015)}]{kunevs2015excitonic}%
  \BibitemOpen
  \bibfield  {author} {\bibinfo {author} {\bibfnamefont {J.}~\bibnamefont {Kune{\v{s}}}},\ }\bibfield  {title} {\bibinfo {title} {Excitonic condensation in systems of strongly correlated electrons},\ }\href@noop {} {\bibfield  {journal} {\bibinfo  {journal} {Journal of Physics: Condensed Matter}\ }\textbf {\bibinfo {volume} {27}},\ \bibinfo {pages} {333201} (\bibinfo {year} {2015})}\BibitemShut {NoStop}%
\bibitem [{\citenamefont {Werner}\ and\ \citenamefont {Murakami}(2020)}]{werner2020nonthermal}%
  \BibitemOpen
  \bibfield  {author} {\bibinfo {author} {\bibfnamefont {P.}~\bibnamefont {Werner}}\ and\ \bibinfo {author} {\bibfnamefont {Y.}~\bibnamefont {Murakami}},\ }\bibfield  {title} {\bibinfo {title} {Nonthermal excitonic condensation near a spin-state transition},\ }\href@noop {} {\bibfield  {journal} {\bibinfo  {journal} {Physical Review B}\ }\textbf {\bibinfo {volume} {102}},\ \bibinfo {pages} {241103} (\bibinfo {year} {2020})}\BibitemShut {NoStop}%
\bibitem [{\citenamefont {Yang}(1989)}]{yang1989eta}%
  \BibitemOpen
  \bibfield  {author} {\bibinfo {author} {\bibfnamefont {C.~N.}\ \bibnamefont {Yang}},\ }\bibfield  {title} {\bibinfo {title} {{$\eta$ pairing and off-diagonal long-range order in a Hubbard model}},\ }\href@noop {} {\bibfield  {journal} {\bibinfo  {journal} {Physical Review Letters}\ }\textbf {\bibinfo {volume} {63}},\ \bibinfo {pages} {2144} (\bibinfo {year} {1989})}\BibitemShut {NoStop}%
\bibitem [{\citenamefont {Rosch}\ \emph {et~al.}(2008)\citenamefont {Rosch}, \citenamefont {Rasch}, \citenamefont {Binz},\ and\ \citenamefont {Vojta}}]{rosch2008metastable}%
  \BibitemOpen
  \bibfield  {author} {\bibinfo {author} {\bibfnamefont {A.}~\bibnamefont {Rosch}}, \bibinfo {author} {\bibfnamefont {D.}~\bibnamefont {Rasch}}, \bibinfo {author} {\bibfnamefont {B.}~\bibnamefont {Binz}},\ and\ \bibinfo {author} {\bibfnamefont {M.}~\bibnamefont {Vojta}},\ }\bibfield  {title} {\bibinfo {title} {Metastable superfluidity of repulsive fermionic atoms in optical lattices},\ }\href@noop {} {\bibfield  {journal} {\bibinfo  {journal} {Physical Review Letters}\ }\textbf {\bibinfo {volume} {101}},\ \bibinfo {pages} {265301} (\bibinfo {year} {2008})}\BibitemShut {NoStop}%
\bibitem [{\citenamefont {Li}\ \emph {et~al.}(2020{\natexlab{a}})\citenamefont {Li}, \citenamefont {Golez}, \citenamefont {Werner},\ and\ \citenamefont {Eckstein}}]{li2020eta}%
  \BibitemOpen
  \bibfield  {author} {\bibinfo {author} {\bibfnamefont {J.}~\bibnamefont {Li}}, \bibinfo {author} {\bibfnamefont {D.}~\bibnamefont {Golez}}, \bibinfo {author} {\bibfnamefont {P.}~\bibnamefont {Werner}},\ and\ \bibinfo {author} {\bibfnamefont {M.}~\bibnamefont {Eckstein}},\ }\bibfield  {title} {\bibinfo {title} {{$\eta$-paired superconducting hidden phase in photodoped Mott insulators}},\ }\href@noop {} {\bibfield  {journal} {\bibinfo  {journal} {Physical Review B}\ }\textbf {\bibinfo {volume} {102}},\ \bibinfo {pages} {165136} (\bibinfo {year} {2020}{\natexlab{a}})}\BibitemShut {NoStop}%
\bibitem [{\citenamefont {Padma}\ \emph {et~al.}(2025{\natexlab{a}})\citenamefont {Padma}, \citenamefont {Glerean}, \citenamefont {TenHuisen}, \citenamefont {Shen}, \citenamefont {Wang}, \citenamefont {Xu}, \citenamefont {Elliott}, \citenamefont {Homes}, \citenamefont {Skoropata}, \citenamefont {Ueda}, \citenamefont {Liu}, \citenamefont {Paris}, \citenamefont {Romaguera}, \citenamefont {Lee}, \citenamefont {He}, \citenamefont {Wang}, \citenamefont {Lee}, \citenamefont {Choi}, \citenamefont {Park}, \citenamefont {Mao}, \citenamefont {Calandra}, \citenamefont {Jang}, \citenamefont {Razzoli}, \citenamefont {Dean}, \citenamefont {Wang},\ and\ \citenamefont {Mitrano}}]{padma2025symmetry}%
  \BibitemOpen
  \bibfield  {author} {\bibinfo {author} {\bibfnamefont {H.}~\bibnamefont {Padma}}, \bibinfo {author} {\bibfnamefont {F.}~\bibnamefont {Glerean}}, \bibinfo {author} {\bibfnamefont {S.~F.~R.}\ \bibnamefont {TenHuisen}}, \bibinfo {author} {\bibfnamefont {Z.}~\bibnamefont {Shen}}, \bibinfo {author} {\bibfnamefont {H.}~\bibnamefont {Wang}}, \bibinfo {author} {\bibfnamefont {L.}~\bibnamefont {Xu}}, \bibinfo {author} {\bibfnamefont {J.~D.}\ \bibnamefont {Elliott}}, \bibinfo {author} {\bibfnamefont {C.~C.}\ \bibnamefont {Homes}}, \bibinfo {author} {\bibfnamefont {E.}~\bibnamefont {Skoropata}}, \bibinfo {author} {\bibfnamefont {H.}~\bibnamefont {Ueda}}, \bibinfo {author} {\bibfnamefont {B.}~\bibnamefont {Liu}}, \bibinfo {author} {\bibfnamefont {E.}~\bibnamefont {Paris}}, \bibinfo {author} {\bibfnamefont {A.}~\bibnamefont {Romaguera}}, \bibinfo {author} {\bibfnamefont {B.}~\bibnamefont {Lee}}, \bibinfo {author} {\bibfnamefont {W.}~\bibnamefont {He}}, \bibinfo {author} {\bibfnamefont {Y.}~\bibnamefont {Wang}}, \bibinfo
  {author} {\bibfnamefont {S.~H.}\ \bibnamefont {Lee}}, \bibinfo {author} {\bibfnamefont {H.}~\bibnamefont {Choi}}, \bibinfo {author} {\bibfnamefont {S.-Y.}\ \bibnamefont {Park}}, \bibinfo {author} {\bibfnamefont {Z.}~\bibnamefont {Mao}}, \bibinfo {author} {\bibfnamefont {M.}~\bibnamefont {Calandra}}, \bibinfo {author} {\bibfnamefont {H.}~\bibnamefont {Jang}}, \bibinfo {author} {\bibfnamefont {E.}~\bibnamefont {Razzoli}}, \bibinfo {author} {\bibfnamefont {M.~P.~M.}\ \bibnamefont {Dean}}, \bibinfo {author} {\bibfnamefont {Y.}~\bibnamefont {Wang}},\ and\ \bibinfo {author} {\bibfnamefont {M.}~\bibnamefont {Mitrano}},\ }\bibfield  {title} {\bibinfo {title} {{Symmetry-protected electronic metastability in an optically driven cuprate ladder}},\ }\href {https://doi.org/10.1038/s41563-025-02254-2} {\bibfield  {journal} {\bibinfo  {journal} {Nature Materials}\ }\textbf {\bibinfo {volume} {24}},\ \bibinfo {pages} {1584} (\bibinfo {year} {2025}{\natexlab{a}})}\BibitemShut {NoStop}%
\bibitem [{\citenamefont {Abbamonte}\ \emph {et~al.}(2004)\citenamefont {Abbamonte}, \citenamefont {Blumberg}, \citenamefont {Rusydi}, \citenamefont {Gozar}, \citenamefont {Evans}, \citenamefont {Siegrist}, \citenamefont {Venema}, \citenamefont {Eisaki}, \citenamefont {Isaacs},\ and\ \citenamefont {Sawatzky}}]{abbamonte2004crystallization}%
  \BibitemOpen
  \bibfield  {author} {\bibinfo {author} {\bibfnamefont {P.}~\bibnamefont {Abbamonte}}, \bibinfo {author} {\bibfnamefont {G.}~\bibnamefont {Blumberg}}, \bibinfo {author} {\bibfnamefont {A.}~\bibnamefont {Rusydi}}, \bibinfo {author} {\bibfnamefont {A.}~\bibnamefont {Gozar}}, \bibinfo {author} {\bibfnamefont {P.}~\bibnamefont {Evans}}, \bibinfo {author} {\bibfnamefont {T.}~\bibnamefont {Siegrist}}, \bibinfo {author} {\bibfnamefont {L.}~\bibnamefont {Venema}}, \bibinfo {author} {\bibfnamefont {H.}~\bibnamefont {Eisaki}}, \bibinfo {author} {\bibfnamefont {E.}~\bibnamefont {Isaacs}},\ and\ \bibinfo {author} {\bibfnamefont {G.}~\bibnamefont {Sawatzky}},\ }\bibfield  {title} {\bibinfo {title} {Crystallization of charge holes in the spin ladder of {Sr}$_{14}${Cu}$_{24}${O}$_{41}$},\ }\href@noop {} {\bibfield  {journal} {\bibinfo  {journal} {Nature}\ }\textbf {\bibinfo {volume} {431}},\ \bibinfo {pages} {1078} (\bibinfo {year} {2004})}\BibitemShut {NoStop}%
\bibitem [{\citenamefont {Vuleti{\'c}}\ \emph {et~al.}(2006)\citenamefont {Vuleti{\'c}}, \citenamefont {Korin-Hamzi{\'c}}, \citenamefont {Ivek}, \citenamefont {Tomi{\'c}}, \citenamefont {Gorshunov}, \citenamefont {Dressel},\ and\ \citenamefont {Akimitsu}}]{Vuletic2006spinladder}%
  \BibitemOpen
  \bibfield  {author} {\bibinfo {author} {\bibfnamefont {T.}~\bibnamefont {Vuleti{\'c}}}, \bibinfo {author} {\bibfnamefont {B.}~\bibnamefont {Korin-Hamzi{\'c}}}, \bibinfo {author} {\bibfnamefont {T.}~\bibnamefont {Ivek}}, \bibinfo {author} {\bibfnamefont {S.}~\bibnamefont {Tomi{\'c}}}, \bibinfo {author} {\bibfnamefont {B.}~\bibnamefont {Gorshunov}}, \bibinfo {author} {\bibfnamefont {M.}~\bibnamefont {Dressel}},\ and\ \bibinfo {author} {\bibfnamefont {J.}~\bibnamefont {Akimitsu}},\ }\bibfield  {title} {\bibinfo {title} {The spin-ladder and spin-chain system {(La,Y,Sr,Ca)$_{14}$Cu$_{24}$O$_{41}$}: Electronic phases, charge and spin dynamics},\ }\href {https://doi.org/https://doi.org/10.1016/j.physrep.2006.01.005} {\bibfield  {journal} {\bibinfo  {journal} {Physics Reports}\ }\textbf {\bibinfo {volume} {428}},\ \bibinfo {pages} {169} (\bibinfo {year} {2006})}\BibitemShut {NoStop}%
\bibitem [{\citenamefont {Uehara}\ \emph {et~al.}(1996)\citenamefont {Uehara}, \citenamefont {Nagata}, \citenamefont {Akimitsu}, \citenamefont {Takahashi}, \citenamefont {M{\^o}ri},\ and\ \citenamefont {Kinoshita}}]{uehara1996superconductivity}%
  \BibitemOpen
  \bibfield  {author} {\bibinfo {author} {\bibfnamefont {M.}~\bibnamefont {Uehara}}, \bibinfo {author} {\bibfnamefont {T.}~\bibnamefont {Nagata}}, \bibinfo {author} {\bibfnamefont {J.}~\bibnamefont {Akimitsu}}, \bibinfo {author} {\bibfnamefont {H.}~\bibnamefont {Takahashi}}, \bibinfo {author} {\bibfnamefont {N.}~\bibnamefont {M{\^o}ri}},\ and\ \bibinfo {author} {\bibfnamefont {K.}~\bibnamefont {Kinoshita}},\ }\bibfield  {title} {\bibinfo {title} {Superconductivity in the ladder material {Sr}$_{0.4}${Ca}$_{13. 6}${Cu}$_{24}${O}$_{41. 84}$},\ }\href@noop {} {\bibfield  {journal} {\bibinfo  {journal} {Journal of the Physical Society of Japan}\ }\textbf {\bibinfo {volume} {65}},\ \bibinfo {pages} {2764} (\bibinfo {year} {1996})}\BibitemShut {NoStop}%
\bibitem [{\citenamefont {Dagotto}(1999)}]{dagotto1999experiments}%
  \BibitemOpen
  \bibfield  {author} {\bibinfo {author} {\bibfnamefont {E.}~\bibnamefont {Dagotto}},\ }\bibfield  {title} {\bibinfo {title} {Experiments on ladders reveal a complex interplay between a spin-gapped normal state and superconductivity},\ }\href@noop {} {\bibfield  {journal} {\bibinfo  {journal} {Reports on Progress in Physics}\ }\textbf {\bibinfo {volume} {62}},\ \bibinfo {pages} {1525} (\bibinfo {year} {1999})}\BibitemShut {NoStop}%
\bibitem [{\citenamefont {Hirthe}\ \emph {et~al.}(2023)\citenamefont {Hirthe}, \citenamefont {Chalopin}, \citenamefont {Bourgund}, \citenamefont {Bojovi{\'c}}, \citenamefont {Bohrdt}, \citenamefont {Demler}, \citenamefont {Grusdt}, \citenamefont {Bloch},\ and\ \citenamefont {Hilker}}]{hirthe2023magnetically}%
  \BibitemOpen
  \bibfield  {author} {\bibinfo {author} {\bibfnamefont {S.}~\bibnamefont {Hirthe}}, \bibinfo {author} {\bibfnamefont {T.}~\bibnamefont {Chalopin}}, \bibinfo {author} {\bibfnamefont {D.}~\bibnamefont {Bourgund}}, \bibinfo {author} {\bibfnamefont {P.}~\bibnamefont {Bojovi{\'c}}}, \bibinfo {author} {\bibfnamefont {A.}~\bibnamefont {Bohrdt}}, \bibinfo {author} {\bibfnamefont {E.}~\bibnamefont {Demler}}, \bibinfo {author} {\bibfnamefont {F.}~\bibnamefont {Grusdt}}, \bibinfo {author} {\bibfnamefont {I.}~\bibnamefont {Bloch}},\ and\ \bibinfo {author} {\bibfnamefont {T.~A.}\ \bibnamefont {Hilker}},\ }\bibfield  {title} {\bibinfo {title} {Magnetically mediated hole pairing in fermionic ladders of ultracold atoms},\ }\href@noop {} {\bibfield  {journal} {\bibinfo  {journal} {Nature}\ }\textbf {\bibinfo {volume} {613}},\ \bibinfo {pages} {463} (\bibinfo {year} {2023})}\BibitemShut {NoStop}%
\bibitem [{\citenamefont {Padma}\ \emph {et~al.}(2025{\natexlab{b}})\citenamefont {Padma}, \citenamefont {Thomas}, \citenamefont {TenHuisen}, \citenamefont {He}, \citenamefont {Guan}, \citenamefont {Li}, \citenamefont {Lee}, \citenamefont {Wang}, \citenamefont {Lee}, \citenamefont {Mao}, \citenamefont {Jang}, \citenamefont {Bisogni}, \citenamefont {Pelliciari}, \citenamefont {Dean}, \citenamefont {Johnston},\ and\ \citenamefont {Mitrano}}]{padma2025beyond}%
  \BibitemOpen
  \bibfield  {author} {\bibinfo {author} {\bibfnamefont {H.}~\bibnamefont {Padma}}, \bibinfo {author} {\bibfnamefont {J.}~\bibnamefont {Thomas}}, \bibinfo {author} {\bibfnamefont {S.~F.~R.}\ \bibnamefont {TenHuisen}}, \bibinfo {author} {\bibfnamefont {W.}~\bibnamefont {He}}, \bibinfo {author} {\bibfnamefont {Z.}~\bibnamefont {Guan}}, \bibinfo {author} {\bibfnamefont {J.}~\bibnamefont {Li}}, \bibinfo {author} {\bibfnamefont {B.}~\bibnamefont {Lee}}, \bibinfo {author} {\bibfnamefont {Y.}~\bibnamefont {Wang}}, \bibinfo {author} {\bibfnamefont {S.~H.}\ \bibnamefont {Lee}}, \bibinfo {author} {\bibfnamefont {Z.}~\bibnamefont {Mao}}, \bibinfo {author} {\bibfnamefont {H.}~\bibnamefont {Jang}}, \bibinfo {author} {\bibfnamefont {V.}~\bibnamefont {Bisogni}}, \bibinfo {author} {\bibfnamefont {J.}~\bibnamefont {Pelliciari}}, \bibinfo {author} {\bibfnamefont {M.~P.~M.}\ \bibnamefont {Dean}}, \bibinfo {author} {\bibfnamefont {S.}~\bibnamefont {Johnston}},\ and\ \bibinfo {author} {\bibfnamefont {M.}~\bibnamefont {Mitrano}},\
  }\bibfield  {title} {\bibinfo {title} {{Beyond-Hubbard Pairing in a Cuprate Ladder}},\ }\href {https://doi.org/10.1103/PhysRevX.15.021049} {\bibfield  {journal} {\bibinfo  {journal} {Phys. Rev. X}\ }\textbf {\bibinfo {volume} {15}},\ \bibinfo {pages} {021049} (\bibinfo {year} {2025}{\natexlab{b}})}\BibitemShut {NoStop}%
\bibitem [{\citenamefont {Scheie}\ \emph {et~al.}(2025)\citenamefont {Scheie}, \citenamefont {Laurell}, \citenamefont {Thomas}, \citenamefont {Sharma}, \citenamefont {Kolesnikov}, \citenamefont {Granroth}, \citenamefont {Zhang}, \citenamefont {Lake}, \citenamefont {Mihalik~Jr}, \citenamefont {Bewley} \emph {et~al.}}]{scheie2025cooper}%
  \BibitemOpen
  \bibfield  {author} {\bibinfo {author} {\bibfnamefont {A.}~\bibnamefont {Scheie}}, \bibinfo {author} {\bibfnamefont {P.}~\bibnamefont {Laurell}}, \bibinfo {author} {\bibfnamefont {J.}~\bibnamefont {Thomas}}, \bibinfo {author} {\bibfnamefont {V.}~\bibnamefont {Sharma}}, \bibinfo {author} {\bibfnamefont {A.}~\bibnamefont {Kolesnikov}}, \bibinfo {author} {\bibfnamefont {G.}~\bibnamefont {Granroth}}, \bibinfo {author} {\bibfnamefont {Q.}~\bibnamefont {Zhang}}, \bibinfo {author} {\bibfnamefont {B.}~\bibnamefont {Lake}}, \bibinfo {author} {\bibfnamefont {M.}~\bibnamefont {Mihalik~Jr}}, \bibinfo {author} {\bibfnamefont {R.}~\bibnamefont {Bewley}}, \emph {et~al.},\ }\bibfield  {title} {\bibinfo {title} {Cooper-pair localization in the magnetic dynamics of a cuprate ladder},\ }\href@noop {} {\bibfield  {journal} {\bibinfo  {journal} {arXiv preprint arXiv:2501.10296}\ } (\bibinfo {year} {2025})}\BibitemShut {NoStop}%
\bibitem [{swi(2025)}]{swissfel_furka}%
  \BibitemOpen
  \href@noop {} {\bibinfo {title} {{SwissFEL Furka}}},\ \bibinfo {howpublished} {\url{https://www.psi.ch/en/swissfel/furka}} (\bibinfo {year} {2025}),\ \bibinfo {note} {accessed: 23 April 2025}\BibitemShut {NoStop}%
\bibitem [{\citenamefont {N{\"u}cker}\ \emph {et~al.}(2000)\citenamefont {N{\"u}cker}, \citenamefont {Merz}, \citenamefont {Kuntscher}, \citenamefont {Gerhold}, \citenamefont {Schuppler}, \citenamefont {Neudert}, \citenamefont {Golden}, \citenamefont {Fink}, \citenamefont {Schild}, \citenamefont {Stadler} \emph {et~al.}}]{nucker2000hole}%
  \BibitemOpen
  \bibfield  {author} {\bibinfo {author} {\bibfnamefont {N.}~\bibnamefont {N{\"u}cker}}, \bibinfo {author} {\bibfnamefont {M.}~\bibnamefont {Merz}}, \bibinfo {author} {\bibfnamefont {C.~A.}\ \bibnamefont {Kuntscher}}, \bibinfo {author} {\bibfnamefont {S.}~\bibnamefont {Gerhold}}, \bibinfo {author} {\bibfnamefont {S.}~\bibnamefont {Schuppler}}, \bibinfo {author} {\bibfnamefont {R.}~\bibnamefont {Neudert}}, \bibinfo {author} {\bibfnamefont {M.}~\bibnamefont {Golden}}, \bibinfo {author} {\bibfnamefont {J.}~\bibnamefont {Fink}}, \bibinfo {author} {\bibfnamefont {D.}~\bibnamefont {Schild}}, \bibinfo {author} {\bibfnamefont {S.}~\bibnamefont {Stadler}}, \emph {et~al.},\ }\bibfield  {title} {\bibinfo {title} {Hole distribution in {(Sr,Ca,Y,La)}$_{14}${Cu}$_{24}${O}$_{41}$ ladder compounds studied by x-ray absorption spectroscopy},\ }\href@noop {} {\bibfield  {journal} {\bibinfo  {journal} {Physical Review B}\ }\textbf {\bibinfo {volume} {62}},\ \bibinfo {pages} {14384} (\bibinfo {year} {2000})}\BibitemShut {NoStop}%
\bibitem [{\citenamefont {Mitrano}\ and\ \citenamefont {Wang}(2020)}]{mitrano2020probing}%
  \BibitemOpen
  \bibfield  {author} {\bibinfo {author} {\bibfnamefont {M.}~\bibnamefont {Mitrano}}\ and\ \bibinfo {author} {\bibfnamefont {Y.}~\bibnamefont {Wang}},\ }\bibfield  {title} {\bibinfo {title} {{Probing light-driven quantum materials with ultrafast resonant inelastic X-ray scattering}},\ }\href {https://doi.org/10.1038/s42005-020-00447-6} {\bibfield  {journal} {\bibinfo  {journal} {Communications Physics}\ }\textbf {\bibinfo {volume} {3}},\ \bibinfo {pages} {184} (\bibinfo {year} {2020})}\BibitemShut {NoStop}%
\bibitem [{\citenamefont {Mitrano}\ \emph {et~al.}(2024)\citenamefont {Mitrano}, \citenamefont {Johnston}, \citenamefont {Kim},\ and\ \citenamefont {Dean}}]{Mitrano2024exploring}%
  \BibitemOpen
  \bibfield  {author} {\bibinfo {author} {\bibfnamefont {M.}~\bibnamefont {Mitrano}}, \bibinfo {author} {\bibfnamefont {S.}~\bibnamefont {Johnston}}, \bibinfo {author} {\bibfnamefont {Y.-J.}\ \bibnamefont {Kim}},\ and\ \bibinfo {author} {\bibfnamefont {M.~P.~M.}\ \bibnamefont {Dean}},\ }\bibfield  {title} {\bibinfo {title} {{Exploring Quantum Materials with Resonant Inelastic X-Ray Scattering}},\ }\href {https://doi.org/10.1103/PhysRevX.14.040501} {\bibfield  {journal} {\bibinfo  {journal} {Phys. Rev. X}\ }\textbf {\bibinfo {volume} {14}},\ \bibinfo {pages} {040501} (\bibinfo {year} {2024})}\BibitemShut {NoStop}%
\bibitem [{\citenamefont {Rusydi}\ \emph {et~al.}(2006)\citenamefont {Rusydi}, \citenamefont {Abbamonte}, \citenamefont {Eisaki}, \citenamefont {Fujimaki}, \citenamefont {Blumberg}, \citenamefont {Uchida},\ and\ \citenamefont {Sawatzky}}]{Rusydi2006quantum}%
  \BibitemOpen
  \bibfield  {author} {\bibinfo {author} {\bibfnamefont {A.}~\bibnamefont {Rusydi}}, \bibinfo {author} {\bibfnamefont {P.}~\bibnamefont {Abbamonte}}, \bibinfo {author} {\bibfnamefont {H.}~\bibnamefont {Eisaki}}, \bibinfo {author} {\bibfnamefont {Y.}~\bibnamefont {Fujimaki}}, \bibinfo {author} {\bibfnamefont {G.}~\bibnamefont {Blumberg}}, \bibinfo {author} {\bibfnamefont {S.}~\bibnamefont {Uchida}},\ and\ \bibinfo {author} {\bibfnamefont {G.~A.}\ \bibnamefont {Sawatzky}},\ }\bibfield  {title} {\bibinfo {title} {Quantum melting of the hole crystal in the spin ladder of {Sr}$_{14-x}${Ca}$_x${Cu}$_{24}${O}$_{41}$},\ }\href {https://doi.org/10.1103/PhysRevLett.97.016403} {\bibfield  {journal} {\bibinfo  {journal} {Physical Review Letters}\ }\textbf {\bibinfo {volume} {97}},\ \bibinfo {pages} {016403} (\bibinfo {year} {2006})}\BibitemShut {NoStop}%
\bibitem [{\citenamefont {Tseng}\ \emph {et~al.}(2023)\citenamefont {Tseng}, \citenamefont {Paris}, \citenamefont {Schmidt}, \citenamefont {Zhang}, \citenamefont {Asmara}, \citenamefont {Bag}, \citenamefont {Strocov}, \citenamefont {Singh}, \citenamefont {Schlappa}, \citenamefont {R{\o}nnow} \emph {et~al.}}]{tseng2023momentum}%
  \BibitemOpen
  \bibfield  {author} {\bibinfo {author} {\bibfnamefont {Y.}~\bibnamefont {Tseng}}, \bibinfo {author} {\bibfnamefont {E.}~\bibnamefont {Paris}}, \bibinfo {author} {\bibfnamefont {K.~P.}\ \bibnamefont {Schmidt}}, \bibinfo {author} {\bibfnamefont {W.}~\bibnamefont {Zhang}}, \bibinfo {author} {\bibfnamefont {T.~C.}\ \bibnamefont {Asmara}}, \bibinfo {author} {\bibfnamefont {R.}~\bibnamefont {Bag}}, \bibinfo {author} {\bibfnamefont {V.~N.}\ \bibnamefont {Strocov}}, \bibinfo {author} {\bibfnamefont {S.}~\bibnamefont {Singh}}, \bibinfo {author} {\bibfnamefont {J.}~\bibnamefont {Schlappa}}, \bibinfo {author} {\bibfnamefont {H.~M.}\ \bibnamefont {R{\o}nnow}}, \emph {et~al.},\ }\bibfield  {title} {\bibinfo {title} {Momentum-resolved spin-conserving two-triplon bound state and continuum in a cuprate ladder},\ }\href@noop {} {\bibfield  {journal} {\bibinfo  {journal} {Communications Physics}\ }\textbf {\bibinfo {volume} {6}},\ \bibinfo {pages} {138} (\bibinfo {year} {2023})}\BibitemShut {NoStop}%
\bibitem [{\citenamefont {Lee}\ \emph {et~al.}(2014)\citenamefont {Lee}, \citenamefont {Lee}, \citenamefont {Nowadnick}, \citenamefont {Gerber}, \citenamefont {Tabis}, \citenamefont {Huang}, \citenamefont {Strocov}, \citenamefont {Motoyama}, \citenamefont {Yu}, \citenamefont {Moritz} \emph {et~al.}}]{lee2014asymmetry}%
  \BibitemOpen
  \bibfield  {author} {\bibinfo {author} {\bibfnamefont {W.}~\bibnamefont {Lee}}, \bibinfo {author} {\bibfnamefont {J.}~\bibnamefont {Lee}}, \bibinfo {author} {\bibfnamefont {E.}~\bibnamefont {Nowadnick}}, \bibinfo {author} {\bibfnamefont {S.}~\bibnamefont {Gerber}}, \bibinfo {author} {\bibfnamefont {W.}~\bibnamefont {Tabis}}, \bibinfo {author} {\bibfnamefont {S.}~\bibnamefont {Huang}}, \bibinfo {author} {\bibfnamefont {V.}~\bibnamefont {Strocov}}, \bibinfo {author} {\bibfnamefont {E.}~\bibnamefont {Motoyama}}, \bibinfo {author} {\bibfnamefont {G.}~\bibnamefont {Yu}}, \bibinfo {author} {\bibfnamefont {B.}~\bibnamefont {Moritz}}, \emph {et~al.},\ }\bibfield  {title} {\bibinfo {title} {Asymmetry of collective excitations in electron-and hole-doped cuprate superconductors},\ }\href@noop {} {\bibfield  {journal} {\bibinfo  {journal} {Nature Physics}\ }\textbf {\bibinfo {volume} {10}},\ \bibinfo {pages} {883} (\bibinfo {year} {2014})}\BibitemShut {NoStop}%
\bibitem [{\citenamefont {Hepting}\ \emph {et~al.}(2018)\citenamefont {Hepting}, \citenamefont {Chaix}, \citenamefont {Huang}, \citenamefont {Fumagalli}, \citenamefont {Peng}, \citenamefont {Moritz}, \citenamefont {Kummer}, \citenamefont {Brookes}, \citenamefont {Lee}, \citenamefont {Hashimoto} \emph {et~al.}}]{hepting2018three}%
  \BibitemOpen
  \bibfield  {author} {\bibinfo {author} {\bibfnamefont {M.}~\bibnamefont {Hepting}}, \bibinfo {author} {\bibfnamefont {L.}~\bibnamefont {Chaix}}, \bibinfo {author} {\bibfnamefont {E.}~\bibnamefont {Huang}}, \bibinfo {author} {\bibfnamefont {R.}~\bibnamefont {Fumagalli}}, \bibinfo {author} {\bibfnamefont {Y.}~\bibnamefont {Peng}}, \bibinfo {author} {\bibfnamefont {B.}~\bibnamefont {Moritz}}, \bibinfo {author} {\bibfnamefont {K.}~\bibnamefont {Kummer}}, \bibinfo {author} {\bibfnamefont {N.}~\bibnamefont {Brookes}}, \bibinfo {author} {\bibfnamefont {W.}~\bibnamefont {Lee}}, \bibinfo {author} {\bibfnamefont {M.}~\bibnamefont {Hashimoto}}, \emph {et~al.},\ }\bibfield  {title} {\bibinfo {title} {Three-dimensional collective charge excitations in electron-doped copper oxide superconductors},\ }\href@noop {} {\bibfield  {journal} {\bibinfo  {journal} {Nature}\ }\textbf {\bibinfo {volume} {563}},\ \bibinfo {pages} {374} (\bibinfo {year} {2018})}\BibitemShut {NoStop}%
\bibitem [{\citenamefont {{Lin}}\ \emph {et~al.}(2020)\citenamefont {{Lin}}, \citenamefont {{Yuan}}, \citenamefont {{Jin}}, \citenamefont {{Yin}}, \citenamefont {{Li}}, \citenamefont {{Zhou}}, \citenamefont {{Lu}}, \citenamefont {{Dantz}}, \citenamefont {{Schmitt}}, \citenamefont {{Ding}}, \citenamefont {{Guo}}, \citenamefont {{Dean}},\ and\ \citenamefont {{Liu}}}]{lin2020doping}%
  \BibitemOpen
  \bibfield  {author} {\bibinfo {author} {\bibfnamefont {J.~Q.}\ \bibnamefont {{Lin}}}, \bibinfo {author} {\bibfnamefont {J.}~\bibnamefont {{Yuan}}}, \bibinfo {author} {\bibfnamefont {K.}~\bibnamefont {{Jin}}}, \bibinfo {author} {\bibfnamefont {Z.~P.}\ \bibnamefont {{Yin}}}, \bibinfo {author} {\bibfnamefont {G.}~\bibnamefont {{Li}}}, \bibinfo {author} {\bibfnamefont {K.-J.}\ \bibnamefont {{Zhou}}}, \bibinfo {author} {\bibfnamefont {X.}~\bibnamefont {{Lu}}}, \bibinfo {author} {\bibfnamefont {M.}~\bibnamefont {{Dantz}}}, \bibinfo {author} {\bibfnamefont {T.}~\bibnamefont {{Schmitt}}}, \bibinfo {author} {\bibfnamefont {H.}~\bibnamefont {{Ding}}}, \bibinfo {author} {\bibfnamefont {H.}~\bibnamefont {{Guo}}}, \bibinfo {author} {\bibfnamefont {M.~P.~M.}\ \bibnamefont {{Dean}}},\ and\ \bibinfo {author} {\bibfnamefont {X.}~\bibnamefont {{Liu}}},\ }\bibfield  {title} {\bibinfo {title} {{Doping evolution of the charge excitations and electron correlations in electron-doped superconducting La${2-x}$Ce$_x$CuO$_4$}},\
  }\href {https://doi.org/10.1038/s41535-019-0205-9} {\bibfield  {journal} {\bibinfo  {journal} {npj Quantum Materials}\ }\textbf {\bibinfo {volume} {5}},\ \bibinfo {pages} {4} (\bibinfo {year} {2020})}\BibitemShut {NoStop}%
\bibitem [{\citenamefont {Nag}\ \emph {et~al.}(2020)\citenamefont {Nag}, \citenamefont {Zhu}, \citenamefont {Bejas}, \citenamefont {Li}, \citenamefont {Robarts}, \citenamefont {Yamase}, \citenamefont {Petsch}, \citenamefont {Song}, \citenamefont {Eisaki}, \citenamefont {Walters} \emph {et~al.}}]{nag2020detection}%
  \BibitemOpen
  \bibfield  {author} {\bibinfo {author} {\bibfnamefont {A.}~\bibnamefont {Nag}}, \bibinfo {author} {\bibfnamefont {M.}~\bibnamefont {Zhu}}, \bibinfo {author} {\bibfnamefont {M.}~\bibnamefont {Bejas}}, \bibinfo {author} {\bibfnamefont {J.}~\bibnamefont {Li}}, \bibinfo {author} {\bibfnamefont {H.}~\bibnamefont {Robarts}}, \bibinfo {author} {\bibfnamefont {H.}~\bibnamefont {Yamase}}, \bibinfo {author} {\bibfnamefont {A.}~\bibnamefont {Petsch}}, \bibinfo {author} {\bibfnamefont {D.}~\bibnamefont {Song}}, \bibinfo {author} {\bibfnamefont {H.}~\bibnamefont {Eisaki}}, \bibinfo {author} {\bibfnamefont {A.}~\bibnamefont {Walters}}, \emph {et~al.},\ }\bibfield  {title} {\bibinfo {title} {{Detection of acoustic plasmons in hole-doped lanthanum and bismuth cuprate superconductors using resonant inelastic X-ray scattering}},\ }\href@noop {} {\bibfield  {journal} {\bibinfo  {journal} {Physical Review Letters}\ }\textbf {\bibinfo {volume} {125}},\ \bibinfo {pages} {257002} (\bibinfo {year} {2020})}\BibitemShut {NoStop}%
\bibitem [{\citenamefont {Lomeli}\ \emph {et~al.}(2025)\citenamefont {Lomeli}, \citenamefont {Kundu}, \citenamefont {Chuang}, \citenamefont {Zhuo}, \citenamefont {Chen}, \citenamefont {Xi}, \citenamefont {Shen}, \citenamefont {Dakovski}, \citenamefont {Gepr{\"a}gs}, \citenamefont {Moritz} \emph {et~al.}}]{lomeli2025direct}%
  \BibitemOpen
  \bibfield  {author} {\bibinfo {author} {\bibfnamefont {E.~G.}\ \bibnamefont {Lomeli}}, \bibinfo {author} {\bibfnamefont {S.}~\bibnamefont {Kundu}}, \bibinfo {author} {\bibfnamefont {Y.-D.}\ \bibnamefont {Chuang}}, \bibinfo {author} {\bibfnamefont {Z.}~\bibnamefont {Zhuo}}, \bibinfo {author} {\bibfnamefont {K.}~\bibnamefont {Chen}}, \bibinfo {author} {\bibfnamefont {X.}~\bibnamefont {Xi}}, \bibinfo {author} {\bibfnamefont {L.}~\bibnamefont {Shen}}, \bibinfo {author} {\bibfnamefont {G.~L.}\ \bibnamefont {Dakovski}}, \bibinfo {author} {\bibfnamefont {S.}~\bibnamefont {Gepr{\"a}gs}}, \bibinfo {author} {\bibfnamefont {B.}~\bibnamefont {Moritz}}, \emph {et~al.},\ }\bibfield  {title} {\bibinfo {title} {Direct observation of the {Lindhard} continuum using resonant inelastic x-ray scattering},\ }\href@noop {} {\bibfield  {journal} {\bibinfo  {journal} {arXiv preprint arXiv:2509.10741}\ } (\bibinfo {year} {2025})}\BibitemShut {NoStop}%
\bibitem [{\citenamefont {Giamarchi}(2003)}]{giamarchi2003quantum}%
  \BibitemOpen
  \bibfield  {author} {\bibinfo {author} {\bibfnamefont {T.}~\bibnamefont {Giamarchi}},\ }\href@noop {} {\emph {\bibinfo {title} {Quantum physics in one dimension}}},\ Vol.\ \bibinfo {volume} {121}\ (\bibinfo  {publisher} {Clarendon press},\ \bibinfo {year} {2003})\BibitemShut {NoStop}%
\bibitem [{\citenamefont {Kung}\ \emph {et~al.}(2017)\citenamefont {Kung}, \citenamefont {Bazin}, \citenamefont {Wohlfeld}, \citenamefont {Wang}, \citenamefont {Chen}, \citenamefont {Jia}, \citenamefont {Johnston}, \citenamefont {Moritz}, \citenamefont {Mila},\ and\ \citenamefont {Devereaux}}]{kung2017numerically}%
  \BibitemOpen
  \bibfield  {author} {\bibinfo {author} {\bibfnamefont {Y.}~\bibnamefont {Kung}}, \bibinfo {author} {\bibfnamefont {C.}~\bibnamefont {Bazin}}, \bibinfo {author} {\bibfnamefont {K.}~\bibnamefont {Wohlfeld}}, \bibinfo {author} {\bibfnamefont {Y.}~\bibnamefont {Wang}}, \bibinfo {author} {\bibfnamefont {C.-C.}\ \bibnamefont {Chen}}, \bibinfo {author} {\bibfnamefont {C.}~\bibnamefont {Jia}}, \bibinfo {author} {\bibfnamefont {S.}~\bibnamefont {Johnston}}, \bibinfo {author} {\bibfnamefont {B.}~\bibnamefont {Moritz}}, \bibinfo {author} {\bibfnamefont {F.}~\bibnamefont {Mila}},\ and\ \bibinfo {author} {\bibfnamefont {T.}~\bibnamefont {Devereaux}},\ }\bibfield  {title} {\bibinfo {title} {Numerically exploring the {1D-2D} dimensional crossover on spin dynamics in the doped {Hubbard} model},\ }\href@noop {} {\bibfield  {journal} {\bibinfo  {journal} {Physical Review B}\ }\textbf {\bibinfo {volume} {96}},\ \bibinfo {pages} {195106} (\bibinfo {year} {2017})}\BibitemShut {NoStop}%
\bibitem [{\citenamefont {Kivelson}\ \emph {et~al.}(2003)\citenamefont {Kivelson}, \citenamefont {Bindloss}, \citenamefont {Fradkin}, \citenamefont {Oganesyan}, \citenamefont {Tranquada}, \citenamefont {Kapitulnik},\ and\ \citenamefont {Howald}}]{kivelson2003detect}%
  \BibitemOpen
  \bibfield  {author} {\bibinfo {author} {\bibfnamefont {S.~A.}\ \bibnamefont {Kivelson}}, \bibinfo {author} {\bibfnamefont {I.~P.}\ \bibnamefont {Bindloss}}, \bibinfo {author} {\bibfnamefont {E.}~\bibnamefont {Fradkin}}, \bibinfo {author} {\bibfnamefont {V.}~\bibnamefont {Oganesyan}}, \bibinfo {author} {\bibfnamefont {J.}~\bibnamefont {Tranquada}}, \bibinfo {author} {\bibfnamefont {A.}~\bibnamefont {Kapitulnik}},\ and\ \bibinfo {author} {\bibfnamefont {C.}~\bibnamefont {Howald}},\ }\bibfield  {title} {\bibinfo {title} {How to detect fluctuating stripes in the high-temperature superconductors},\ }\href@noop {} {\bibfield  {journal} {\bibinfo  {journal} {Reviews of Modern Physics}\ }\textbf {\bibinfo {volume} {75}},\ \bibinfo {pages} {1201} (\bibinfo {year} {2003})}\BibitemShut {NoStop}%
\bibitem [{\citenamefont {Mitrano}\ \emph {et~al.}(2019{\natexlab{a}})\citenamefont {Mitrano}, \citenamefont {Lee}, \citenamefont {Husain}, \citenamefont {Delacretaz}, \citenamefont {Zhu}, \citenamefont {de~la Pe{\~n}a~Munoz}, \citenamefont {Sun}, \citenamefont {Joe}, \citenamefont {Reid}, \citenamefont {Wandel} \emph {et~al.}}]{mitrano2019ultrafast}%
  \BibitemOpen
  \bibfield  {author} {\bibinfo {author} {\bibfnamefont {M.}~\bibnamefont {Mitrano}}, \bibinfo {author} {\bibfnamefont {S.}~\bibnamefont {Lee}}, \bibinfo {author} {\bibfnamefont {A.~A.}\ \bibnamefont {Husain}}, \bibinfo {author} {\bibfnamefont {L.}~\bibnamefont {Delacretaz}}, \bibinfo {author} {\bibfnamefont {M.}~\bibnamefont {Zhu}}, \bibinfo {author} {\bibfnamefont {G.}~\bibnamefont {de~la Pe{\~n}a~Munoz}}, \bibinfo {author} {\bibfnamefont {S.~X.-L.}\ \bibnamefont {Sun}}, \bibinfo {author} {\bibfnamefont {Y.~I.}\ \bibnamefont {Joe}}, \bibinfo {author} {\bibfnamefont {A.~H.}\ \bibnamefont {Reid}}, \bibinfo {author} {\bibfnamefont {S.~F.}\ \bibnamefont {Wandel}}, \emph {et~al.},\ }\bibfield  {title} {\bibinfo {title} {{Ultrafast time-resolved x-ray scattering reveals diffusive charge order dynamics in La$_{2-x}$Ba$_x$CuO$_4$}},\ }\href@noop {} {\bibfield  {journal} {\bibinfo  {journal} {Science Advances}\ }\textbf {\bibinfo {volume} {5}},\ \bibinfo {pages} {eaax3346} (\bibinfo {year}
  {2019}{\natexlab{a}})}\BibitemShut {NoStop}%
\bibitem [{\citenamefont {Mitrano}\ \emph {et~al.}(2019{\natexlab{b}})\citenamefont {Mitrano}, \citenamefont {Lee}, \citenamefont {Husain}, \citenamefont {Zhu}, \citenamefont {Munoz}, \citenamefont {Sun}, \citenamefont {Joe}, \citenamefont {Reid}, \citenamefont {Wandel}, \citenamefont {Coslovich}, \citenamefont {Schlotter}, \citenamefont {van Driel}, \citenamefont {Schneeloch}, \citenamefont {Gu}, \citenamefont {Goldenfeld},\ and\ \citenamefont {Abbamonte}}]{Mitrano2019evidence}%
  \BibitemOpen
  \bibfield  {author} {\bibinfo {author} {\bibfnamefont {M.}~\bibnamefont {Mitrano}}, \bibinfo {author} {\bibfnamefont {S.}~\bibnamefont {Lee}}, \bibinfo {author} {\bibfnamefont {A.~A.}\ \bibnamefont {Husain}}, \bibinfo {author} {\bibfnamefont {M.}~\bibnamefont {Zhu}}, \bibinfo {author} {\bibfnamefont {G.~d. l. P.~n.}\ \bibnamefont {Munoz}}, \bibinfo {author} {\bibfnamefont {S.~X.-L.}\ \bibnamefont {Sun}}, \bibinfo {author} {\bibfnamefont {Y.~I.}\ \bibnamefont {Joe}}, \bibinfo {author} {\bibfnamefont {A.~H.}\ \bibnamefont {Reid}}, \bibinfo {author} {\bibfnamefont {S.~F.}\ \bibnamefont {Wandel}}, \bibinfo {author} {\bibfnamefont {G.}~\bibnamefont {Coslovich}}, \bibinfo {author} {\bibfnamefont {W.}~\bibnamefont {Schlotter}}, \bibinfo {author} {\bibfnamefont {T.}~\bibnamefont {van Driel}}, \bibinfo {author} {\bibfnamefont {J.}~\bibnamefont {Schneeloch}}, \bibinfo {author} {\bibfnamefont {G.~D.}\ \bibnamefont {Gu}}, \bibinfo {author} {\bibfnamefont {N.}~\bibnamefont {Goldenfeld}},\ and\ \bibinfo {author}
  {\bibfnamefont {P.}~\bibnamefont {Abbamonte}},\ }\bibfield  {title} {\bibinfo {title} {{Evidence for photoinduced sliding of the charge-order condensate in La$_{1.875}$Ba$_{0.125}$CuO$_4$}},\ }\href {https://doi.org/10.1103/PhysRevB.100.205125} {\bibfield  {journal} {\bibinfo  {journal} {Physical Review B}\ }\textbf {\bibinfo {volume} {100}},\ \bibinfo {pages} {205125} (\bibinfo {year} {2019}{\natexlab{b}})}\BibitemShut {NoStop}%
\bibitem [{\citenamefont {Chaix}\ \emph {et~al.}(2017)\citenamefont {Chaix}, \citenamefont {Ghiringhelli}, \citenamefont {Peng}, \citenamefont {Hashimoto}, \citenamefont {Moritz}, \citenamefont {Kummer}, \citenamefont {Brookes}, \citenamefont {He}, \citenamefont {Chen}, \citenamefont {Ishida} \emph {et~al.}}]{chaix2017dispersive}%
  \BibitemOpen
  \bibfield  {author} {\bibinfo {author} {\bibfnamefont {L.}~\bibnamefont {Chaix}}, \bibinfo {author} {\bibfnamefont {G.}~\bibnamefont {Ghiringhelli}}, \bibinfo {author} {\bibfnamefont {Y.}~\bibnamefont {Peng}}, \bibinfo {author} {\bibfnamefont {M.}~\bibnamefont {Hashimoto}}, \bibinfo {author} {\bibfnamefont {B.}~\bibnamefont {Moritz}}, \bibinfo {author} {\bibfnamefont {K.}~\bibnamefont {Kummer}}, \bibinfo {author} {\bibfnamefont {N.~B.}\ \bibnamefont {Brookes}}, \bibinfo {author} {\bibfnamefont {Y.}~\bibnamefont {He}}, \bibinfo {author} {\bibfnamefont {S.}~\bibnamefont {Chen}}, \bibinfo {author} {\bibfnamefont {S.}~\bibnamefont {Ishida}}, \emph {et~al.},\ }\bibfield  {title} {\bibinfo {title} {Dispersive charge density wave excitations in {Bi}$_2${Sr}$_2${Ca}{Cu}$_2${O}$_{8+\delta}$},\ }\href@noop {} {\bibfield  {journal} {\bibinfo  {journal} {Nature Physics}\ }\textbf {\bibinfo {volume} {13}},\ \bibinfo {pages} {952} (\bibinfo {year} {2017})}\BibitemShut {NoStop}%
\bibitem [{\citenamefont {Lee}\ \emph {et~al.}(2021)\citenamefont {Lee}, \citenamefont {Zhou}, \citenamefont {Hepting}, \citenamefont {Li}, \citenamefont {Nag}, \citenamefont {Walters}, \citenamefont {Garcia-Fernandez}, \citenamefont {Robarts}, \citenamefont {Hashimoto}, \citenamefont {Lu} \emph {et~al.}}]{lee2021spectroscopic}%
  \BibitemOpen
  \bibfield  {author} {\bibinfo {author} {\bibfnamefont {W.-S.}\ \bibnamefont {Lee}}, \bibinfo {author} {\bibfnamefont {K.-J.}\ \bibnamefont {Zhou}}, \bibinfo {author} {\bibfnamefont {M.}~\bibnamefont {Hepting}}, \bibinfo {author} {\bibfnamefont {J.}~\bibnamefont {Li}}, \bibinfo {author} {\bibfnamefont {A.}~\bibnamefont {Nag}}, \bibinfo {author} {\bibfnamefont {A.}~\bibnamefont {Walters}}, \bibinfo {author} {\bibfnamefont {M.}~\bibnamefont {Garcia-Fernandez}}, \bibinfo {author} {\bibfnamefont {H.}~\bibnamefont {Robarts}}, \bibinfo {author} {\bibfnamefont {M.}~\bibnamefont {Hashimoto}}, \bibinfo {author} {\bibfnamefont {H.}~\bibnamefont {Lu}}, \emph {et~al.},\ }\bibfield  {title} {\bibinfo {title} {Spectroscopic fingerprint of charge order melting driven by quantum fluctuations in a cuprate},\ }\href@noop {} {\bibfield  {journal} {\bibinfo  {journal} {Nature Physics}\ }\textbf {\bibinfo {volume} {17}},\ \bibinfo {pages} {53} (\bibinfo {year} {2021})}\BibitemShut {NoStop}%
\bibitem [{\citenamefont {Lin}\ \emph {et~al.}(2020)\citenamefont {Lin}, \citenamefont {Miao}, \citenamefont {Mazzone}, \citenamefont {Gu}, \citenamefont {Nag}, \citenamefont {Walters}, \citenamefont {Garc{\'\i}a-Fern{\'a}ndez}, \citenamefont {Barbour}, \citenamefont {Pelliciari}, \citenamefont {Jarrige} \emph {et~al.}}]{lin2020strongly}%
  \BibitemOpen
  \bibfield  {author} {\bibinfo {author} {\bibfnamefont {J.}~\bibnamefont {Lin}}, \bibinfo {author} {\bibfnamefont {H.}~\bibnamefont {Miao}}, \bibinfo {author} {\bibfnamefont {D.}~\bibnamefont {Mazzone}}, \bibinfo {author} {\bibfnamefont {G.}~\bibnamefont {Gu}}, \bibinfo {author} {\bibfnamefont {A.}~\bibnamefont {Nag}}, \bibinfo {author} {\bibfnamefont {A.}~\bibnamefont {Walters}}, \bibinfo {author} {\bibfnamefont {M.}~\bibnamefont {Garc{\'\i}a-Fern{\'a}ndez}}, \bibinfo {author} {\bibfnamefont {A.}~\bibnamefont {Barbour}}, \bibinfo {author} {\bibfnamefont {J.}~\bibnamefont {Pelliciari}}, \bibinfo {author} {\bibfnamefont {I.}~\bibnamefont {Jarrige}}, \emph {et~al.},\ }\bibfield  {title} {\bibinfo {title} {Strongly correlated charge density wave in {La}$_{2-x}${Sr}$_x${Cu}{O}$_4$ evidenced by doping-dependent phonon anomaly},\ }\href@noop {} {\bibfield  {journal} {\bibinfo  {journal} {Physical Review Letters}\ }\textbf {\bibinfo {volume} {124}},\ \bibinfo {pages} {207005} (\bibinfo {year} {2020})}\BibitemShut
  {NoStop}%
\bibitem [{\citenamefont {Li}\ \emph {et~al.}(2020{\natexlab{b}})\citenamefont {Li}, \citenamefont {Nag}, \citenamefont {Pelliciari}, \citenamefont {Robarts}, \citenamefont {Walters}, \citenamefont {Garcia-Fernandez}, \citenamefont {Eisaki}, \citenamefont {Song}, \citenamefont {Ding}, \citenamefont {Johnston} \emph {et~al.}}]{li2020multiorbital}%
  \BibitemOpen
  \bibfield  {author} {\bibinfo {author} {\bibfnamefont {J.}~\bibnamefont {Li}}, \bibinfo {author} {\bibfnamefont {A.}~\bibnamefont {Nag}}, \bibinfo {author} {\bibfnamefont {J.}~\bibnamefont {Pelliciari}}, \bibinfo {author} {\bibfnamefont {H.}~\bibnamefont {Robarts}}, \bibinfo {author} {\bibfnamefont {A.}~\bibnamefont {Walters}}, \bibinfo {author} {\bibfnamefont {M.}~\bibnamefont {Garcia-Fernandez}}, \bibinfo {author} {\bibfnamefont {H.}~\bibnamefont {Eisaki}}, \bibinfo {author} {\bibfnamefont {D.}~\bibnamefont {Song}}, \bibinfo {author} {\bibfnamefont {H.}~\bibnamefont {Ding}}, \bibinfo {author} {\bibfnamefont {S.}~\bibnamefont {Johnston}}, \emph {et~al.},\ }\bibfield  {title} {\bibinfo {title} {Multiorbital charge-density wave excitations and concomitant phonon anomalies in {Bi}$_2${Sr}$_2${La}{Cu}{O}$_{6+ \delta}$},\ }\href@noop {} {\bibfield  {journal} {\bibinfo  {journal} {Proceedings of the National Academy of Sciences}\ }\textbf {\bibinfo {volume} {117}},\ \bibinfo {pages} {16219} (\bibinfo {year}
  {2020}{\natexlab{b}})}\BibitemShut {NoStop}%
\bibitem [{\citenamefont {Huang}\ \emph {et~al.}(2021)\citenamefont {Huang}, \citenamefont {Singh}, \citenamefont {Mou}, \citenamefont {Johnston}, \citenamefont {Kemper}, \citenamefont {van~den Brink}, \citenamefont {Chen}, \citenamefont {Lee}, \citenamefont {Okamoto}, \citenamefont {Chu} \emph {et~al.}}]{huang2021quantum}%
  \BibitemOpen
  \bibfield  {author} {\bibinfo {author} {\bibfnamefont {H.}~\bibnamefont {Huang}}, \bibinfo {author} {\bibfnamefont {A.}~\bibnamefont {Singh}}, \bibinfo {author} {\bibfnamefont {C.}~\bibnamefont {Mou}}, \bibinfo {author} {\bibfnamefont {S.}~\bibnamefont {Johnston}}, \bibinfo {author} {\bibfnamefont {A.}~\bibnamefont {Kemper}}, \bibinfo {author} {\bibfnamefont {J.}~\bibnamefont {van~den Brink}}, \bibinfo {author} {\bibfnamefont {P.}~\bibnamefont {Chen}}, \bibinfo {author} {\bibfnamefont {T.}~\bibnamefont {Lee}}, \bibinfo {author} {\bibfnamefont {J.}~\bibnamefont {Okamoto}}, \bibinfo {author} {\bibfnamefont {Y.}~\bibnamefont {Chu}}, \emph {et~al.},\ }\bibfield  {title} {\bibinfo {title} {Quantum fluctuations of charge order induce phonon softening in a superconducting cuprate},\ }\href@noop {} {\bibfield  {journal} {\bibinfo  {journal} {Physical Review X}\ }\textbf {\bibinfo {volume} {11}},\ \bibinfo {pages} {041038} (\bibinfo {year} {2021})}\BibitemShut {NoStop}%
\bibitem [{\citenamefont {Takahashi}\ \emph {et~al.}(1997)\citenamefont {Takahashi}, \citenamefont {Yokoya}, \citenamefont {Ashihara}, \citenamefont {Akaki}, \citenamefont {Fujisawa}, \citenamefont {Chainani}, \citenamefont {Uehara}, \citenamefont {Nagata}, \citenamefont {Akimitsu},\ and\ \citenamefont {Tsunetsugu}}]{takahashi1997angle}%
  \BibitemOpen
  \bibfield  {author} {\bibinfo {author} {\bibfnamefont {T.}~\bibnamefont {Takahashi}}, \bibinfo {author} {\bibfnamefont {T.}~\bibnamefont {Yokoya}}, \bibinfo {author} {\bibfnamefont {A.}~\bibnamefont {Ashihara}}, \bibinfo {author} {\bibfnamefont {O.}~\bibnamefont {Akaki}}, \bibinfo {author} {\bibfnamefont {H.}~\bibnamefont {Fujisawa}}, \bibinfo {author} {\bibfnamefont {A.}~\bibnamefont {Chainani}}, \bibinfo {author} {\bibfnamefont {M.}~\bibnamefont {Uehara}}, \bibinfo {author} {\bibfnamefont {T.}~\bibnamefont {Nagata}}, \bibinfo {author} {\bibfnamefont {J.}~\bibnamefont {Akimitsu}},\ and\ \bibinfo {author} {\bibfnamefont {H.}~\bibnamefont {Tsunetsugu}},\ }\bibfield  {title} {\bibinfo {title} {Angle-resolved photoemission study of the ladder compound {Sr}$_{14}${Cu}$_{24}${O}$_{41}$},\ }\href@noop {} {\bibfield  {journal} {\bibinfo  {journal} {Physical Review B}\ }\textbf {\bibinfo {volume} {56}},\ \bibinfo {pages} {7870} (\bibinfo {year} {1997})}\BibitemShut {NoStop}%
\bibitem [{\citenamefont {Wohlfeld}\ \emph {et~al.}(2007)\citenamefont {Wohlfeld}, \citenamefont {Ole{\'s}},\ and\ \citenamefont {Sawatzky}}]{wohlfeld2007origin}%
  \BibitemOpen
  \bibfield  {author} {\bibinfo {author} {\bibfnamefont {K.}~\bibnamefont {Wohlfeld}}, \bibinfo {author} {\bibfnamefont {A.~M.}\ \bibnamefont {Ole{\'s}}},\ and\ \bibinfo {author} {\bibfnamefont {G.~A.}\ \bibnamefont {Sawatzky}},\ }\bibfield  {title} {\bibinfo {title} {Origin of the charge density wave in coupled spin ladders in {Sr}$_{14-x}${Ca}$_x${Cu}$_{24}${O}$_{41}$},\ }\href@noop {} {\bibfield  {journal} {\bibinfo  {journal} {Physical Review B}\ }\textbf {\bibinfo {volume} {75}},\ \bibinfo {pages} {180501} (\bibinfo {year} {2007})}\BibitemShut {NoStop}%
\bibitem [{\citenamefont {Wohlfeld}\ \emph {et~al.}(2010)\citenamefont {Wohlfeld}, \citenamefont {Ole{\'s}},\ and\ \citenamefont {Sawatzky}}]{wohlfeld2010t}%
  \BibitemOpen
  \bibfield  {author} {\bibinfo {author} {\bibfnamefont {K.}~\bibnamefont {Wohlfeld}}, \bibinfo {author} {\bibfnamefont {A.~M.}\ \bibnamefont {Ole{\'s}}},\ and\ \bibinfo {author} {\bibfnamefont {G.~A.}\ \bibnamefont {Sawatzky}},\ }\bibfield  {title} {\bibinfo {title} {{{t-J} model of coupled {Cu}$_2${O}$_5$ ladders in Sr$_{14-x}$Ca$_x$Cu$_{24}$O$_{41}$}},\ }\href@noop {} {\bibfield  {journal} {\bibinfo  {journal} {Physical Review B}\ }\textbf {\bibinfo {volume} {81}},\ \bibinfo {pages} {214522} (\bibinfo {year} {2010})}\BibitemShut {NoStop}%
\bibitem [{\citenamefont {Nicoletti}\ \emph {et~al.}(2014)\citenamefont {Nicoletti}, \citenamefont {Casandruc}, \citenamefont {Laplace}, \citenamefont {Khanna}, \citenamefont {Hunt}, \citenamefont {Kaiser}, \citenamefont {Dhesi}, \citenamefont {Gu}, \citenamefont {Hill},\ and\ \citenamefont {Cavalleri}}]{nicoletti2014optically}%
  \BibitemOpen
  \bibfield  {author} {\bibinfo {author} {\bibfnamefont {D.}~\bibnamefont {Nicoletti}}, \bibinfo {author} {\bibfnamefont {E.}~\bibnamefont {Casandruc}}, \bibinfo {author} {\bibfnamefont {Y.}~\bibnamefont {Laplace}}, \bibinfo {author} {\bibfnamefont {V.}~\bibnamefont {Khanna}}, \bibinfo {author} {\bibfnamefont {C.~R.}\ \bibnamefont {Hunt}}, \bibinfo {author} {\bibfnamefont {S.}~\bibnamefont {Kaiser}}, \bibinfo {author} {\bibfnamefont {S.}~\bibnamefont {Dhesi}}, \bibinfo {author} {\bibfnamefont {G.}~\bibnamefont {Gu}}, \bibinfo {author} {\bibfnamefont {J.}~\bibnamefont {Hill}},\ and\ \bibinfo {author} {\bibfnamefont {A.}~\bibnamefont {Cavalleri}},\ }\bibfield  {title} {\bibinfo {title} {Optically induced superconductivity in striped {La}$_{2-x}${Ba}$_x${Cu}{O}$_4$ by polarization-selective excitation in the near infrared},\ }\href@noop {} {\bibfield  {journal} {\bibinfo  {journal} {Physical Review B}\ }\textbf {\bibinfo {volume} {90}},\ \bibinfo {pages} {100503} (\bibinfo {year} {2014})}\BibitemShut {NoStop}%
\bibitem [{\citenamefont {Cremin}\ \emph {et~al.}(2019)\citenamefont {Cremin}, \citenamefont {Zhang}, \citenamefont {Homes}, \citenamefont {Gu}, \citenamefont {Sun}, \citenamefont {Fogler}, \citenamefont {Millis}, \citenamefont {Basov},\ and\ \citenamefont {Averitt}}]{cremin2019photoenhanced}%
  \BibitemOpen
  \bibfield  {author} {\bibinfo {author} {\bibfnamefont {K.~A.}\ \bibnamefont {Cremin}}, \bibinfo {author} {\bibfnamefont {J.}~\bibnamefont {Zhang}}, \bibinfo {author} {\bibfnamefont {C.~C.}\ \bibnamefont {Homes}}, \bibinfo {author} {\bibfnamefont {G.~D.}\ \bibnamefont {Gu}}, \bibinfo {author} {\bibfnamefont {Z.}~\bibnamefont {Sun}}, \bibinfo {author} {\bibfnamefont {M.~M.}\ \bibnamefont {Fogler}}, \bibinfo {author} {\bibfnamefont {A.~J.}\ \bibnamefont {Millis}}, \bibinfo {author} {\bibfnamefont {D.~N.}\ \bibnamefont {Basov}},\ and\ \bibinfo {author} {\bibfnamefont {R.~D.}\ \bibnamefont {Averitt}},\ }\bibfield  {title} {\bibinfo {title} {Photoenhanced metastable c-axis electrodynamics in stripe-ordered cuprate {La}$_{1.885}${Ba}$_{0.115}${Cu}{O}$_4$},\ }\href@noop {} {\bibfield  {journal} {\bibinfo  {journal} {Proceedings of the National Academy of Sciences}\ }\textbf {\bibinfo {volume} {116}},\ \bibinfo {pages} {19875} (\bibinfo {year} {2019})}\BibitemShut {NoStop}%
\bibitem [{\citenamefont {Wang}\ \emph {et~al.}(2021)\citenamefont {Wang}, \citenamefont {Shi},\ and\ \citenamefont {Chen}}]{wang2021fluctuating}%
  \BibitemOpen
  \bibfield  {author} {\bibinfo {author} {\bibfnamefont {Y.}~\bibnamefont {Wang}}, \bibinfo {author} {\bibfnamefont {T.}~\bibnamefont {Shi}},\ and\ \bibinfo {author} {\bibfnamefont {C.-C.}\ \bibnamefont {Chen}},\ }\bibfield  {title} {\bibinfo {title} {Fluctuating nature of light-enhanced d-wave superconductivity: a time-dependent variational non-gaussian exact diagonalization study},\ }\href@noop {} {\bibfield  {journal} {\bibinfo  {journal} {Physical Review X}\ }\textbf {\bibinfo {volume} {11}},\ \bibinfo {pages} {041028} (\bibinfo {year} {2021})}\BibitemShut {NoStop}%
\bibitem [{\citenamefont {Dagotto}\ \emph {et~al.}(1992)\citenamefont {Dagotto}, \citenamefont {Riera},\ and\ \citenamefont {Scalapino}}]{dagotto1992superconductivity}%
  \BibitemOpen
  \bibfield  {author} {\bibinfo {author} {\bibfnamefont {E.}~\bibnamefont {Dagotto}}, \bibinfo {author} {\bibfnamefont {J.}~\bibnamefont {Riera}},\ and\ \bibinfo {author} {\bibfnamefont {D.}~\bibnamefont {Scalapino}},\ }\bibfield  {title} {\bibinfo {title} {Superconductivity in ladders and coupled planes},\ }\href@noop {} {\bibfield  {journal} {\bibinfo  {journal} {Physical Review B}\ }\textbf {\bibinfo {volume} {45}},\ \bibinfo {pages} {5744} (\bibinfo {year} {1992})}\BibitemShut {NoStop}%
\bibitem [{\citenamefont {Blumberg}\ \emph {et~al.}(2002)\citenamefont {Blumberg}, \citenamefont {Littlewood}, \citenamefont {Gozar}, \citenamefont {Dennis}, \citenamefont {Motoyama}, \citenamefont {Eisaki},\ and\ \citenamefont {Uchida}}]{blumberg2002sliding}%
  \BibitemOpen
  \bibfield  {author} {\bibinfo {author} {\bibfnamefont {G.}~\bibnamefont {Blumberg}}, \bibinfo {author} {\bibfnamefont {P.}~\bibnamefont {Littlewood}}, \bibinfo {author} {\bibfnamefont {A.}~\bibnamefont {Gozar}}, \bibinfo {author} {\bibfnamefont {B.}~\bibnamefont {Dennis}}, \bibinfo {author} {\bibfnamefont {N.}~\bibnamefont {Motoyama}}, \bibinfo {author} {\bibfnamefont {H.}~\bibnamefont {Eisaki}},\ and\ \bibinfo {author} {\bibfnamefont {S.}~\bibnamefont {Uchida}},\ }\bibfield  {title} {\bibinfo {title} {Sliding density wave in {Sr}$_{14}${Cu}$_{24}${O}$_{41}$ ladder compounds},\ }\href@noop {} {\bibfield  {journal} {\bibinfo  {journal} {Science}\ }\textbf {\bibinfo {volume} {297}},\ \bibinfo {pages} {584} (\bibinfo {year} {2002})}\BibitemShut {NoStop}%
\bibitem [{\citenamefont {Vuleti{\'c}}\ \emph {et~al.}(2003)\citenamefont {Vuleti{\'c}}, \citenamefont {Korin-Hamzi{\'c}}, \citenamefont {Tomi{\'c}}, \citenamefont {Gorshunov}, \citenamefont {Haas}, \citenamefont {Room}, \citenamefont {Dressel}, \citenamefont {Akimitsu}, \citenamefont {Sasaki},\ and\ \citenamefont {Nagata}}]{vuletic2003suppression}%
  \BibitemOpen
  \bibfield  {author} {\bibinfo {author} {\bibfnamefont {T.}~\bibnamefont {Vuleti{\'c}}}, \bibinfo {author} {\bibfnamefont {B.}~\bibnamefont {Korin-Hamzi{\'c}}}, \bibinfo {author} {\bibfnamefont {S.}~\bibnamefont {Tomi{\'c}}}, \bibinfo {author} {\bibfnamefont {B.}~\bibnamefont {Gorshunov}}, \bibinfo {author} {\bibfnamefont {P.}~\bibnamefont {Haas}}, \bibinfo {author} {\bibfnamefont {T.}~\bibnamefont {Room}}, \bibinfo {author} {\bibfnamefont {M.}~\bibnamefont {Dressel}}, \bibinfo {author} {\bibfnamefont {J.}~\bibnamefont {Akimitsu}}, \bibinfo {author} {\bibfnamefont {T.}~\bibnamefont {Sasaki}},\ and\ \bibinfo {author} {\bibfnamefont {T.}~\bibnamefont {Nagata}},\ }\bibfield  {title} {\bibinfo {title} {Suppression of the charge-density-wave state in {Sr}$_{14}${Cu}$_{24}${O}$_{41}$ by {Calcium} doping},\ }\href@noop {} {\bibfield  {journal} {\bibinfo  {journal} {Physical Review Letters}\ }\textbf {\bibinfo {volume} {90}},\ \bibinfo {pages} {257002} (\bibinfo {year} {2003})}\BibitemShut {NoStop}%
\bibitem [{\citenamefont {Murakami}\ \emph {et~al.}(2022)\citenamefont {Murakami}, \citenamefont {Takayoshi}, \citenamefont {Kaneko}, \citenamefont {Sun}, \citenamefont {Gole{\v{z}}}, \citenamefont {Millis},\ and\ \citenamefont {Werner}}]{murakami2022exploring}%
  \BibitemOpen
  \bibfield  {author} {\bibinfo {author} {\bibfnamefont {Y.}~\bibnamefont {Murakami}}, \bibinfo {author} {\bibfnamefont {S.}~\bibnamefont {Takayoshi}}, \bibinfo {author} {\bibfnamefont {T.}~\bibnamefont {Kaneko}}, \bibinfo {author} {\bibfnamefont {Z.}~\bibnamefont {Sun}}, \bibinfo {author} {\bibfnamefont {D.}~\bibnamefont {Gole{\v{z}}}}, \bibinfo {author} {\bibfnamefont {A.~J.}\ \bibnamefont {Millis}},\ and\ \bibinfo {author} {\bibfnamefont {P.}~\bibnamefont {Werner}},\ }\bibfield  {title} {\bibinfo {title} {{Exploring nonequilibrium phases of photo-doped Mott insulators with generalized Gibbs ensembles}},\ }\href@noop {} {\bibfield  {journal} {\bibinfo  {journal} {Communications Physics}\ }\textbf {\bibinfo {volume} {5}},\ \bibinfo {pages} {23} (\bibinfo {year} {2022})}\BibitemShut {NoStop}%
\end{thebibliography}%
\end{document}


\title{\textbf{A light-induced charge order mode in a metastable cuprate ladder} 
}%

\author{Hari Padma}
\email{hpadmanabhan@g.harvard.edu}
\affiliation{Department of Physics, Harvard University, Cambridge, MA, USA}

\author{Prakash Sharma}
\affiliation{Department of Chemistry, Emory University, Atlanta, GA, USA}

\author{Sophia F. R. TenHuisen}
\affiliation{Department of Physics, Harvard University, Cambridge, MA, USA}
\affiliation{Department of Applied Physics, Harvard University, Cambridge, MA, USA}

\author{Filippo Glerean}
\email{Present address: Condensed Matter Physics and Materials Science Department, Brookhaven National Laboratory, Upton, NY, USA}
\affiliation{Department of Physics, Harvard University, Cambridge, MA, USA}

\author{Antoine Roll}
\affiliation{Condensed Matter Physics and Materials Science Department, Brookhaven National Laboratory, Upton, NY, USA}

\author{Pan Zhou}
\affiliation{Department of Chemistry, Emory University, Atlanta, GA, USA}

\author{Sarbajaya Kundu}
\affiliation{Department of Physics and Astronomy, University of Notre Dame, Notre Dame, IN, USA}
\affiliation{Stavropoulos Center for Complex Quantum Matter, University of Notre Dame, Notre Dame, IN, USA}

\author{Arnau Romaguera}
\author{Elizabeth Skoropata}
\author{Hiroki Ueda}
\author{Biaolong Liu}
\author{Eugenio Paris}
\affiliation{PSI Center for Photon Science, Paul Scherrer Institute, Villigen, Switzerland}

\author{Yu Wang}
\author{Seng Huat Lee}
\author{Zhiqiang Mao}
\affiliation{Department of Physics, Pennsylvania State University, University Park, PA, USA}
\affiliation{2D Crystal Consortium, Materials Research Institute, Pennsylvania State University, University Park, PA, USA}

\author{Mark P. M. Dean}
\affiliation{Condensed Matter Physics and Materials Science Department, Brookhaven National Laboratory, Upton, NY, USA}

\author{Edwin W. Huang}
\affiliation{Department of Physics and Astronomy, University of Notre Dame, Notre Dame, IN, USA}
\affiliation{Stavropoulos Center for Complex Quantum Matter, University of Notre Dame, Notre Dame, IN, USA}

\author{Elia Razzoli}
\affiliation{PSI Center for Photon Science, Paul Scherrer Institute, Villigen, Switzerland}

\author{Yao Wang}
\affiliation{Department of Chemistry, Emory University, Atlanta, GA, USA}

\author{Matteo Mitrano}
\email{mmitrano@g.harvard.edu}
\affiliation{Department of Physics, Harvard University, Cambridge, MA, USA}

\maketitle
\vspace{-10mm}
\section*{Supplemental Material}





\begin{enumerate}
\item \textbf{trXAS measurements}
\item \textbf{trRIXS measurements}
\item \textbf{RPA calculations}
\item \textbf{DMRG calculations}
\end{enumerate}

\clearpage

\noindent
\subsection{1. trXAS measurements}



The O $K$-edge trXAS spectra in Fig. 1 feature three prominent peaks. Prior work \cite{nucker2000hole} demonstrated that the peaks centered at 528 eV and 528.5 eV consist primarily of contributions from the chain and ladder Zhang-Rice singlets, respectively, while the peak centered at 529.4 eV includes overlapping contributions from the chain and ladder upper Hubbard bands. The transfer of holes from chains to ladders by Ca substitution at equilibrium is empirically associated with a suppression of the UHB, and an enhancement of the ZRS peaks \cite{nucker2000hole} (also see discussion in SI Section 4 of Ref. \cite{padma2025symmetry}). 

We model the trXAS spectra as the sum of three Lorentzians and a linear function
\vspace{-2mm}
\begin{equation}
\label{eq4}
I(\omega) = L(\omega; A_1, \omega_{01}, \sigma_1) + L(\omega; A_2, \omega_{02}, \sigma_2) + L(\omega; A_3, \omega_{03}, \sigma_3) + a\omega + b,
\end{equation}
\vspace{-0.5mm}
where the Lorentzians $L(\omega; A_i, \omega_{0i}, \sigma_i)$ describe chain ($i$ = 1) and ladder ($i$ = 2) Zhang-Rice singlet peaks, and the upper Hubbard band ($i$ = 3). The Lorentzian function is given by
\begin{equation}
L(\omega; A, \omega_{0}, \sigma) = \frac{A}{\pi}\left[\frac{\sigma}{(\omega - \omega_0)^2 + \sigma^2}\right],
\end{equation}
where $A$, $\omega_0$, and $\sigma$ are the amplitude, center energy, and linewidth parameters. The fit parameters are shown in Table \ref{tab1}.
\vspace{-2mm}
\begin{table}[h]
\caption{Fit parameters for XAS spectra shown in Fig. 1(c-f) of the main text.}\label{tab1}
\begin{tabular}{@{}llllllllllll@{}}
\toprule
   & $A_1$ & $\omega_{01}$ (eV) & $\sigma_1$ (eV) & $A_2$  & $\omega_{02}$ (eV)  & $\sigma_2$ (eV) & $A_3$  & $\omega_{03}$ (eV)  & $\sigma_3$ (eV) & $a$ & $b$ (eV) \\
\midrule
Equilibrium~   & 0.7(1) & 527.89(2) & 0.28(3) & 0.5(1)  & 528.45(4)  & 0.34(6) & 1.5(1)  & 529.68(2)  & 0.55(4) & 0.038(6)  & -20(3) \\
$t$ = 0.4 ps~   & 0.8(1) & 527.90(2) & 0.31(3) & 0.6(1)  & 528.46(4)  & 0.37(6) & 1.5(1)  & 529.70(2)  & 1.18(9) & 0.035(6)  & -18(3)  \\
$t$ = 3 ps~   & 0.7(1) & 527.90(2) & 0.29(3) & 0.6(1)  & 528.45(4)  & 0.36(6) & 1.5(1)  & 529.70(2)  & 0.59(4) & 0.037(6)  & -18(3)  \\
$t$ = 101 ps~   & 0.7(1) & 527.89(2) & 0.29(3) & 0.6(2)  & 528.44(4)  & 0.37(7) & 1.5(1)  & 529.70(2)  & 0.58(5) & 0.034(6)  & -18(3)  \\
\botrule
\end{tabular}
\end{table}

We observe an enhanced ladder Zhang-Rice singlet peak amplitude $A_2$, consistent with chain-to-ladder hole transfer by Ca substitution at equilibrium. Furthermore, all fit parameters are unchanged from 3 ps to 101 ps, confirming that the hole transfer is metastable.

\clearpage

\noindent
\subsection{2. trRIXS measurements}

\begin{figure}[h]
    \centering
    \includegraphics[width=0.625\textwidth]{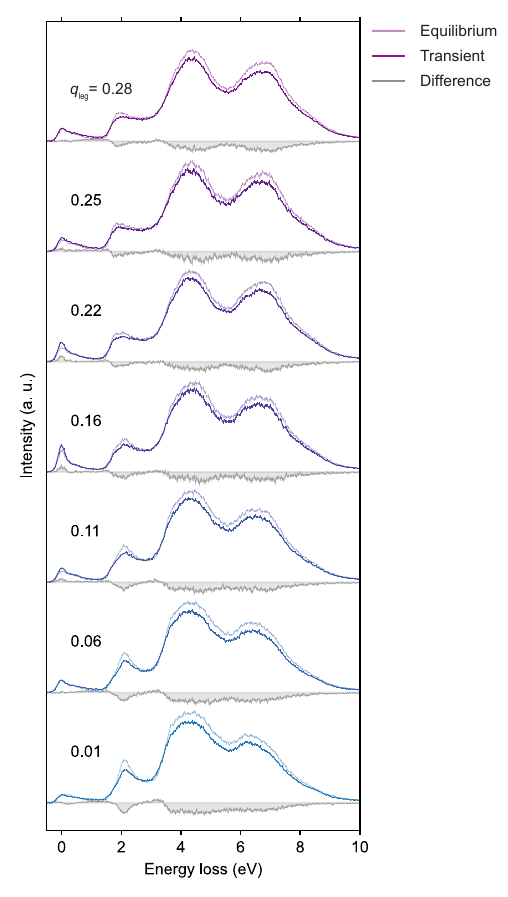}
    \caption{Equilibrium, transient ($t$ = 0.4 ps), and difference trRIXS spectra for momenta spanning $q_\mathrm{leg}$ = 0.01 to 0.28 r.l.u. The spectra are vertically offset for clarity. All spectra are normalized to the equilibrium integrated intensity of $dd$ and charge transfer excitations from 1.3 eV to 10 eV.}
    \label{fig_trRIXS_raw}
\end{figure}

We present the full set of trRIXS spectra in Fig. \ref{fig_trRIXS_raw}. All trRIXS spectra presented in this manuscript are normalized to the equilibrium integrated intensity of $dd$ and charge transfer excitations from 1.3 eV to 10 eV.

\begin{figure}[h]
    \centering
    \includegraphics[width=0.65\textwidth]{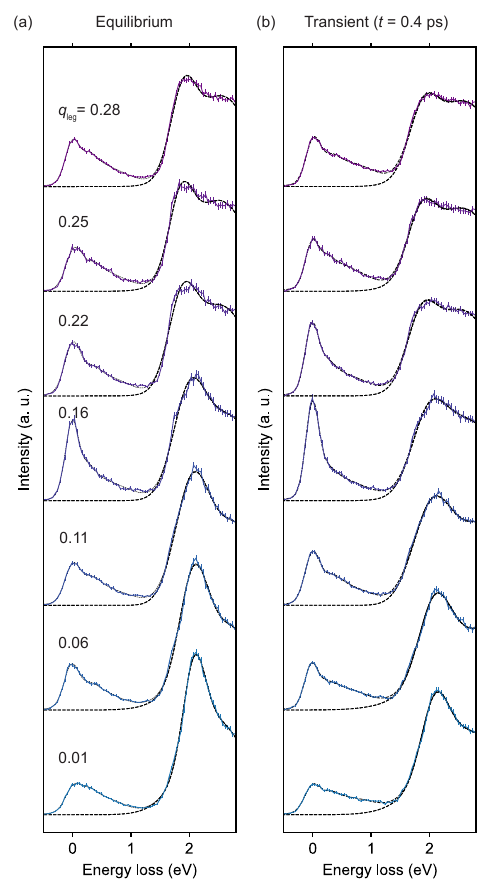}
    \caption{(a) Equilibrium and (b) transient trRIXS spectra for momenta spanning $q_\mathrm{leg}$ = 0.01 to 0.28 r.l.u. The dashed black lines are fits to the $dd$ excitations between 1.3 eV and 2.8 eV. The spectra are vertically offset for clarity.}
    \label{fig_trRIXS_ddfit}
\end{figure}

\begin{figure}[h]
    \centering
    \includegraphics[width=0.45\textwidth]{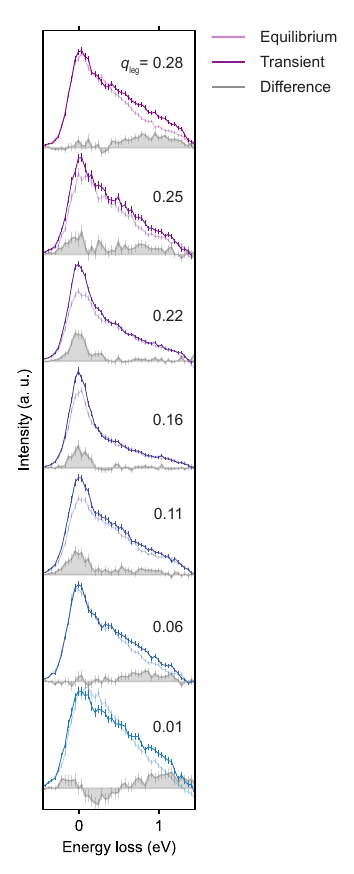}
    \caption{Equilibrium, transient, and difference trRIXS spectra for momenta spanning $q_\mathrm{leg}$ = 0.01 to 0.28 r.l.u., with the $dd$ excitations subtracted using the fits in Fig. \ref{fig_trRIXS_ddfit}. The spectra are vertically offset for clarity.}
    \label{fig_trRIXS_nodd}
\end{figure}

To isolate spectral changes below 1 eV, we fit and subtract the $dd$ excitations above 1.3 eV. We independently fit the equilibrium and transient trRIXS spectra between 1.3 eV and 2.8 eV to the sum of a skewed Gaussian and a Gaussian as shown in Fig. \ref{fig_trRIXS_ddfit}. The resulting spectra with the $dd$ excitations' contribution removed are shown in Fig. \ref{fig_trRIXS_nodd}, which form the basis for all subsequent analysis and interpretation.

We quantify light-induced spectral changes by fitting the momentum-dependent trRIXS spectra. Our prior work \cite{padma2025symmetry} established that hole transfer from chains to ladders disrupts the spin singlets in the ladders, leading to an intensity suppression of the $\Delta S = 1$ two-triplon continuum. We note that this phenomenon occurs without any accompanying energy shifts, suggesting that the exchange constants remain unchanged in the metastable state. We use these prior results to constrain our transient fits. 

\begin{figure}[h]
    \centering
    \includegraphics[width=0.45\textwidth]{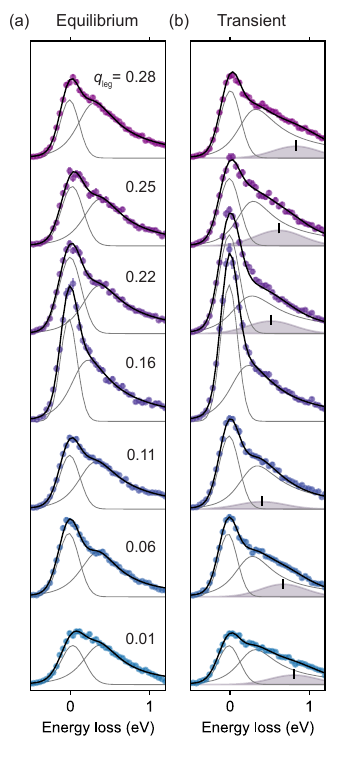}
    \caption{(a) Equilibrium and (b) transient trRIXS spectra for momenta spanning $q_\mathrm{leg}$ = 0.01 to 0.28 r.l.u., fit using the functional form described in the text. Black lines denote the total fits, thin grey lines the individual components. The transient spectra are fit with an additional Gaussian, denoted by grey shaded areas. The spectra are vertically offset for clarity.}
    \label{fig_trRIXS_fits}
\end{figure}

The equilibrium spectra are fit to a Gaussian for the quasielastic peak and an asymmetric Lorentzian (commonly used to fit multimagnon continua, see \cite{bertinshaw2020spin, martinelli2022fractional, padma2025beyond}) for the two-triplon continuum [Fig. \ref{fig_trRIXS_fits}(a)]. For the transient spectra, we include an additional Gaussian to capture the high-energy enhancement. The transient two-triplon frequency, linewidth, and asymmetry are constrained to vary by less than 20\% from their equilibrium values, and the two-triplon amplitude is restricted to not exceed its equilibrium value. Applying this procedure uniformly across all momenta yields the results shown in Fig. \ref{fig_trRIXS_fits}. On average, the $\Delta S = 0$ two-triplon continuum is suppressed by 10\% in the metastable state, similar to the overall suppression of $\Delta S = 1$ channel reported previously \cite{padma2025symmetry}, providing a robust self-consistency check.

\begin{figure}[h]
    \centering
    \includegraphics[width=0.45\textwidth]{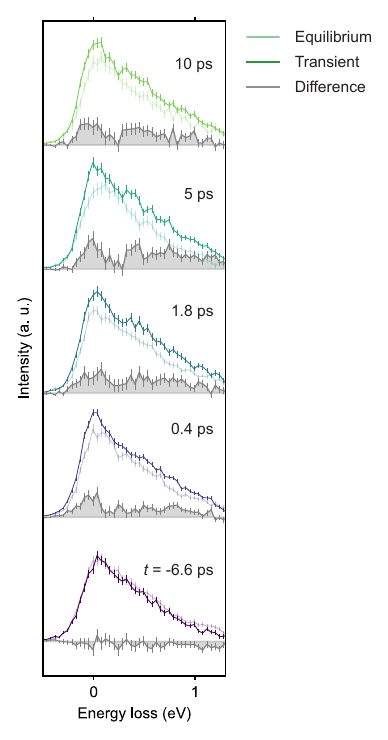}
    \caption{Equilibrium, transient, and difference trRIXS spectra at $q_\mathrm{leg}$ = 0.25 for time delays spanning $t$ = -6.6 to 10 ps. The spectra are vertically offset for clarity.}
    \label{fig_trRIXS_delays}
\end{figure}

To examine the time evolution of the new collective mode, we measure trRIXS spectra as a function of time delay. The results, plotted in Fig. \ref{fig_trRIXS_delays}, show that the transient reshaping of the spectra, including the high-energy enhancement, is unchanged at all positive time delays, indicating that the collective mode is metastable.

\clearpage

\noindent
\subsection{3. RPA calculations}

To check whether plasmons in a system with a charge density wave (CDW) can give rise to modes near $q_\mathrm{CDW}$, we perform a calculation using the random phase approximation (RPA) for a two-leg ladder system. Under the RPA, the loss function is given by
\begin{equation}
    L(\mathbf{q},\omega) = -{\rm Im} \frac{1}{1 - V(\mathbf{q}) \chi_0(\mathbf{q},\omega)}.
\end{equation}

\begin{figure}[h]
    \centering
    \includegraphics[width=0.75\textwidth]{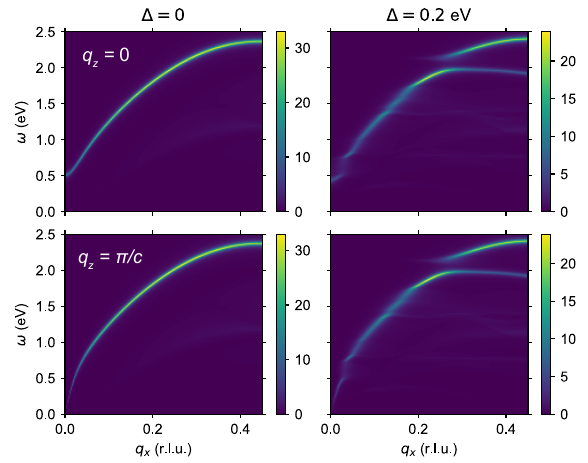}
    \caption{Random phase approximation (RPA) calculations of the loss function, showing expected plasmon behaviors, with or without a CDW order parameter $\Delta$, and at $q_z = 0$ or $q_z = \pi/c$.}
    \label{fig_RPA}
\end{figure}

We use the interaction for a layered electron gas \cite{fetter1974layered,hepting2018three}
\begin{equation}
V(\mathbf{q}) = \frac{V_0}{q_{\|}} \frac{\sinh(q_{\|} c)}{\cosh(q_{\|} c) - \cos(q_z c)},
\end{equation}
where $q_{\|}$ is in-plane momentum transfer, and the Lindhard function
\begin{equation}
    \chi_0(\mathbf{q}, \omega) = \frac{1}{N} \sum_{\mathbf{k}} \frac{f_\mathbf{k} - f_{\mathbf{k}+\mathbf{q}}}{\omega + i0^+ + \epsilon_\mathbf{k} - \epsilon_{\mathbf{k}+\mathbf{q}}}
\end{equation}
where $N=2500$ is the number of $k$-points; $f_\mathbf{k} = [e^{(\epsilon_\mathbf{k}-\mu)/T} + 1]^{-1}$ at temperature $T$, which we set to $0$; and $\epsilon_\mathbf{k}$ is the non-interacting dispersion. We tune the interaction strength $V_0$ to get a plasmon around $0.5$ eV at $q=0$. The results are qualitatively identical for any reasonable choice of $V_0$. As can be seen in the left panels of Fig.\ \ref{fig_RPA}, tuning $q_z$ can turn the plasmon acoustic at small in-plane $q_\|$.

We treat the effects of a CDW at mean-field level by adding to our Hamiltonian a potential
\begin{equation}
    H_{\Delta} = \Delta \sum_{i} \cos\left(\frac{2\pi}{\lambda} x_i\right) \hat{n}_i,
\end{equation}
where $\lambda = 5a$ is the CDW period. The loss function for $\Delta = 0.2$ eV is shown in the right panels of Fig.\ \ref{fig_RPA}. The main effect of the CDW is to break up the plasmon dispersion, due to the more complicated structure of the multi-band Lindhard function resulting in additional Landau damping in some regions. There is no collective mode or ``folding'' associated with the plasmon, and no special behavior near $q_\mathrm{CDW} = 0.2$ r.l.u.

\clearpage

\noindent
\subsection{4. DMRG calculations}
\label{sec:dmrg}

The minimal model for the ladder subunits of Sr$_{14}$Cu$_{24}$O$_{41}$ is the two-leg extended Hubbard model~\cite{padma2025beyond}, described by the Hamiltonian:

\begin{equation}
\begin{aligned}
\hat{H} &= 
     -t \sum_{i,\lambda, \sigma} \left( {c}_{i,\lambda\sigma}^{\dagger} {c}_{i+1,\lambda\sigma} + \mathrm{h.c.} \right) 
     - t_{\perp} \sum_{i,\sigma} \left( {c}_{i,1\sigma}^{\dagger} {c}_{i,2\sigma} + \mathrm{h.c.} \right) \\
    & - t' \sum_{i,\sigma} \left( {c}_{i,1\sigma}^{\dagger} {c}_{i+1,2\sigma} + \mathrm{h.c.} \right) 
     + U \sum_{i,\lambda} {n}_{i,\lambda\uparrow} {n}_{i,\lambda\downarrow} 
     + V \sum_{i,\lambda} {n}_{i\lambda} \hat{n}_{i+1,\lambda}
\end{aligned}
\label{eq:Hamiltonian}
\end{equation}
where \({c}_{i, \lambda\sigma}^{\dagger}\) (\({c}_{i, \lambda\sigma}\)) creates (annihilates) an electron on rung $i$ with chain index $\lambda\in\{1,2\}$ and spin $\sigma\in \{\uparrow,\downarrow\}$, $n_{i,\lambda}= \sum_{\sigma}n_{i, \lambda\sigma}$ is the electron number operator, $t$ and $t_{\perp}$ are the nearest-neighbor hopping strengths along chain and rung directions, respectively, $t'$ is the next-nearest-neighbor (diagonal) hopping between the chains, and $U$ is the onsite Coulomb interaction. Recent studies have revealed that the nearest-neighbor attractive $V\sim-t$ is necessary to accurately model one-dimensional cuprates~\cite{chen2021anomalously, wang2021phonon, padma2025beyond}. 

We adopt the model parameters from previous fits to experimental data~\cite{padma2025beyond}, with $U = 8t$, $t_{\perp} = 0.84t$, and $t' = -0.3t$, where $t \simeq 0.38$ eV for cuprate ladders. Large attractive interactions ($-V > t$) can complicate the computation of long-range correlations due to the proximity to phase separation, hence we fix $V = -0.8t$ in our calculations. We note that our final results are not strongly dependent on the value of $V$. To ensure DMRG convergence, we perform $31$ sweeps and keep the largest singular eigenvalues up to $5600$, which yields a truncation error below $\epsilon = 10^{-6}$ for a ladder of length $L=100$ with open boundary conditions. 

\begin{figure}
    \centering
    \includegraphics[width=0.95\linewidth]{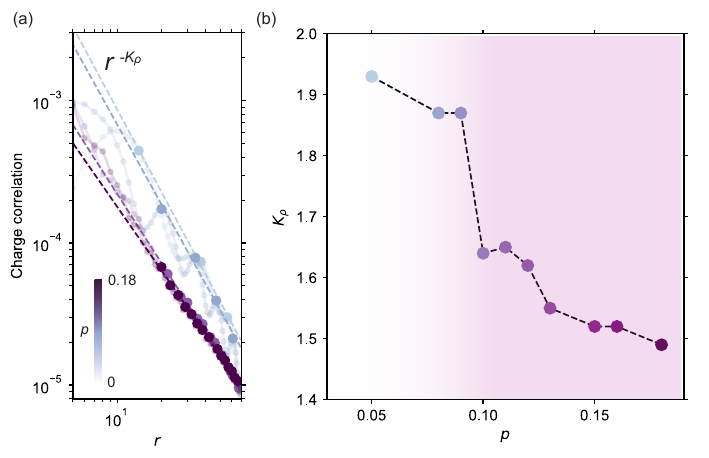}
    \caption{Ground-state DMRG results for two-leg extended Hubbard ladder. (a) The real-space distribution of the charge correlation $D(r)$ at a few representative hole densities $p$, plotted on a logarithmic scale along both axes. The dashed lines are  power law fits, $D(r) \propto r^{-K_\alpha}$. (b) The extracted power law exponents $K_\rho$ as a function of hole density.}
    \label{fig:dmrg_results}
\end{figure}

To analyze the charge properties in the cuprate ladder ground state, we calculate the real-space distribution of charge correlation $D(r)=\langle n_i n_{i+r} \rangle  -\langle n_i \rangle \langle n_{i+r} \rangle$. Due to the sensitivity of the correlation exponents to the choice of the reference bond, we obtain more accurate estimates by averaging over a string of different reference sites in the bulk. We plot $D(r)$ on a logarithmic scale for selected hole densities $p$ in Fig.~\ref{fig:dmrg_results}(a). The charge correlations exhibit a power-law decay at long distances, $D(r) \propto r^{-K_\alpha}$, characterized by the charge exponent $K_\rho$. We plot the extracted exponents as a function of hole density $p$ in Fig.~\ref{fig:dmrg_results}(b). 

The charge correlations exhibit a clear crossover in behavior with increasing doping. At low doping levels, our calculations reveal a charge density wave ground state [see Fig. \ref{fig:dmrg_results}(a)], with highly suppressed long-range charge fluctuations, as indicated by the large exponents $K_\rho$. At larger doping, $K_\rho$ sharply drops at $p \sim 0.1$ [see Fig. \ref{fig:dmrg_results}(b)], together with a quenching of the charge density wave. This sharp drop in $K_\rho$ signifies a rapid increase in the charge susceptibility, supporting the emergence of enhanced charge fluctuations seen experimentally.

\vspace{-6mm}
\bibliography{apssamp}